\documentclass[12pt]{iopart}

\usepackage{iopams}  
\usepackage{color}
\usepackage{graphicx}
\usepackage{soul}
\begin{document}

\title{Preliminary computation of the gap eigenmode of shear Alfv\'{e}n waves on LAPD}

\author{Lei Chang}
\address{School of Aeronautics and Astronautics, Sichuan University, Chengdu 610065, China}
\ead{leichang@scu.edu.cn}

\date{\today}

\begin{abstract}
Characterizing the gap eigenmode of shear Alfv\'{e}n waves (SAW) and its interaction with energetic ions is important to the success of magnetically confined fusion. Previous studies have reported an experimental observation of the spectral gap of SAW on LAPD (Zhang \etal 2008 Phys. Plasmas 15 012103), a linear large plasma device (Gekelman \etal 1991 Rev. Sci. Instrum. 62 2875) possessing easier diagnostic access and lower cost compared with traditional fusion devices, and analytical theory and numerical gap eigenmode using ideal conditions (Chang 2014 PhD Thesis at Australian National University). To guide experimental implementation, the present work models the gap eigenmode of SAW using exact LAPD parameters. A full picture of the wave field for previous experiment reveals that the previously observed spectral gap is not global but an axially local result. To form a global spectral gap, the number of magnetic mirrors has to be increased and stronger static magnetic field makes it more clear. Such a spectral gap is obtained for the magnetic field of $B_0(z)=1.2+0.6\cos[2\pi (z-33.68)/3.63]$ with $7.74$~m magnetic beach. By introducing two types of local defects (corresponding to $E_\theta(z_0)=0$ and $E_\theta'(z_0)=0$ respectively), odd-parity and even-parity discrete eigenmodes are formed clearly inside the gap. The strength of these gap eigenmodes decreases significantly with collision frequency, which is consistent with previous studies. Parameter scans show that these gap eigenmodes can be even formed successfully for the field strength of $B_0(z)=0.2+0.1\cos[2\pi (z-33.68)/3.63]$ and with only $4$ magnetic mirrors, which are achievable by LAPD at its present status. This work can serve as a strong motivation and direct reference for the experimental implementation of the gap eigenmode of SAW on LAPD and other linear plasma devices. 
\end{abstract}

\textbf{Keywords:} Gap eigenmode, shear Alfv\'{e}n wave, LAPD, number and depth of magnetic mirror, linear plasma

\textbf{PACS: 52.35.Bj, 52.55.Jd, 52.25.Xz}

\maketitle

\section{Background}
The gap eigenmode of shear Alfv\'{e}n waves (SAW) has been observed extensively to expel energetic ions from magnetic confinement, which then damage the vessel components of fusion reactor\cite{Duong:1993aa, White:1995aa}. As a result, understanding the formation mechanism of this gap eigenmode and characterizing its interaction with energetic ions are of critical importance to the success of magnetically confined fusion\cite{Cheng:1985aa, Wong:1999aa}. Benefiting from easy diagnostic access, simple geometry and low cost, plasma cylinder with low temperature is a promising candidate to study fundamental physics involved in traditional fusion devices. In 2008, Zhang \etal\cite{Zhang:2008aa} observed a spectral gap in the Alfv\'{e}nic continuum in experiments on LArge Plasma Device (LAPD\cite{Gekelman:1991aa}) by constructing a multiple magnetic mirror array. However, the discrete eigenmode was not formed inside the gap. To guide the experimental implementation of this gap eigenmode, Chang \etal have been developing self-contained analytical theory and carrying out numerical computations first on the gap eigenmode of radially localized helicon waves (RLH)\cite{Chang:2013aa}, which can be excited by energetic electrons in a similar way, and then on the gap eigenmode of SAW for ideal conditions\cite{Chang:2014aa, Chang:2016aa}. The present work computes this gap eigenmode using exact LAPD parameters, including geometry, plasma density, electron temperature, field strength, effective collisionality, and limited number of magnetic mirrors, \etal. It will first give a full picture of the wave field for previous experiment, showing that the previously observed spectral gap is not global but an axially local result, and then clearly form gap eigenmodes for three cases: high field strength and plasma density, low field strength and plasma density, and reduced number of magnetic mirrors. This preliminary computation can be a strong motivation and important reference for the experimental observation of the gap eigenmode of SAW on LAPD and other linear plasma devices.

The paper is organized as following: Sec.~\ref{cmp} describes the numerical scheme and wave field for previous experiment, which also benchmarks our computations; Sec.~\ref{age} presents the spectral gap and gap eigenmode of SAW based on the conditions which are achievable by LAPD either with slight modulations or at its present status; Sec.~\ref{dcl} consists of discussion about continuum damping of this gap eigenmode and conclusions. 

\section{Numerical scheme and wave field for previous experiment }\label{cmp}
This section is devoted to first introducing the numerical scheme and employed conditions, and then showing a full picture of the wave field for previous experiment, which also serves as a detailed benchmark. 

\subsection{Electromagnetic solver (EMS)}
An ElectroMagnetic Solver (EMS)\cite{Chen:2006aa} based on Maxwell's equations and a cold plasma dielectric tensor is employed to study the spectral gap of shear Alfv\'{e}n waves and gap eigenmode inside. The Maxwell's equations are expressed in the frequency domain:
\small
\begin{equation}\label{eq53}
\bigtriangledown\times\mathbf{E}=i \omega\mathbf{B}, 
\end{equation}
\begin{equation}\label{eq54}
\frac{1}{\mu_0}\bigtriangledown\times\mathbf{B}=-i \omega \mathbf{D}+\mathbf{j_a},
\end{equation}
\normalsize
where $\mathbf{E}$ and $\mathbf{B}$ are the wave electric and magnetic fields, respectively, $\mathbf{D}$ is the electric displacement vector, $\mathbf{j_a}$ is the antenna current and $\omega$ is the antenna driving frequency (same to wave frequency for this driven system). These equations are Fourier transformed with respect to the azimuthal angle and then solved (for an azimuthal mode number $m$) by a finite difference scheme on a 2D domain ($r$; $z$). The quantities $\mathbf{D}$ and $\mathbf{E}$ are linked via the cold plasma dielectric tensor\cite{Ginzburg:1964aa}:
\small
\begin{equation}\label{eq55}
\mathbf{D}=\varepsilon_0[\varepsilon\mathbf{E}+ig(\mathbf{E}\times\mathbf{b})+(\eta-\varepsilon)(\mathbf{E}\cdot\mathbf{b})\mathbf{b}],
\end{equation}
\normalsize
where $\mathbf{b}\equiv\mathbf{B_0}/B_0$ is the unit vector along the static magnetic field and 
\small
\begin{equation}\label{eq56}
\begin{array}{l}
\vspace{0.15cm}\varepsilon=1-\sum\limits_{\alpha}\frac{\omega+i\nu_\alpha}{\omega}\frac{\omega^2_{p\alpha}}{(\omega+i\nu_\alpha)^2-\omega^2_{c\alpha}},\\
\vspace{0.15cm}g=-\sum\limits_{\alpha}\frac{\omega_{c\alpha}}{\omega}\frac{\omega^2_{p\alpha}}{(\omega+i\nu_\alpha)^2-\omega^2_{c\alpha}},\\
\eta=1-\sum\limits_{\alpha}\frac{\omega^2_{p\alpha}}{\omega(\omega+i\nu_\alpha)}.
\end{array}
\end{equation}
\normalsize
The subscript $\alpha$ labels particle species (electrons and ions); $\omega_{p\alpha}\equiv\sqrt{n_\alpha q_\alpha^2/\varepsilon_0 m_\alpha}$ is the plasma frequency, $\omega_{c\alpha}\equiv q_\alpha B_0/m_\alpha$ is the cyclotron frequency, and $\nu_\alpha$ a phenomenological collision frequency for each species. The static magnetic field is assumed to be axisymmetric with $B_{0r}~\ll~B_{0z}$ and $B_{0\theta}=0$.\cite{Zhang:2008aa, Chang:2013aa} It is therefore appropriate to use a near axis expansion for the field, namely $B_{0z}$ is only dependent on $z$ and 
\small
\begin{equation}
B_{0r}(r,~z)=-\frac{1}{2}r\frac{\partial B_{0z} (z)}{\partial z}. 
\end{equation}
\normalsize
A blade antenna is employed to excite $m=0$ mode in the plasma, as chosen in the previous experiment\cite{Zhang:2008aa}. The enclosing chamber is assumed to be ideally conducting so that the tangential components of $\mathbf{E}$ vanish at the chamber walls, i. e. 
\small
\begin{equation}
\begin{array}{l}
\vspace{0.2cm}E_\theta(L_r; z)=E_z(L_r; z)=0, \\
\vspace{0.2cm}E_r(r; 0)=E_\theta(r; 0)=0, \\
E_r(r; L_z)=E_\theta(r; L_z)=0,  
\end{array}
\end{equation}
\normalsize
with $L_r$ and $L_z$ the radius and length of the chamber, respectively. Further, all field components must be regular on axis. More details about the EMS can be found in \cite{Chen:2006aa}. 

\subsection{Wave field for previous experiment}
\begin{figure}[ht]
\begin{center}$
\begin{array}{l}
(a)\\
\hspace{0.61cm}\includegraphics[width=0.8\textwidth,angle=0]{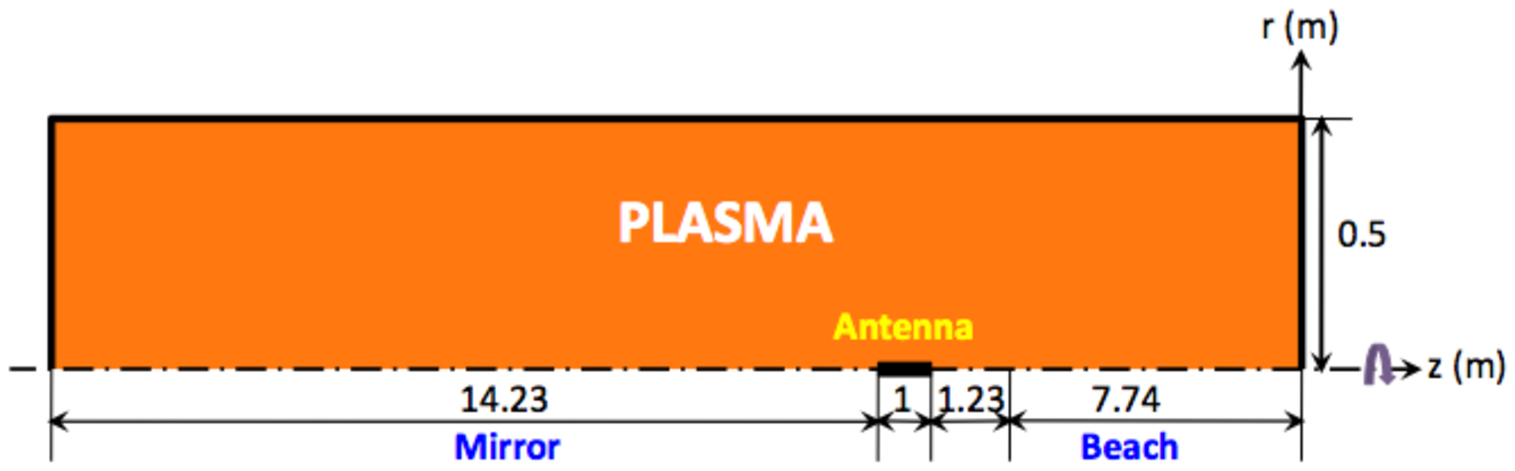}\\
(b)\\
\hspace{-0.07cm}\includegraphics[width=0.735\textwidth,angle=0]{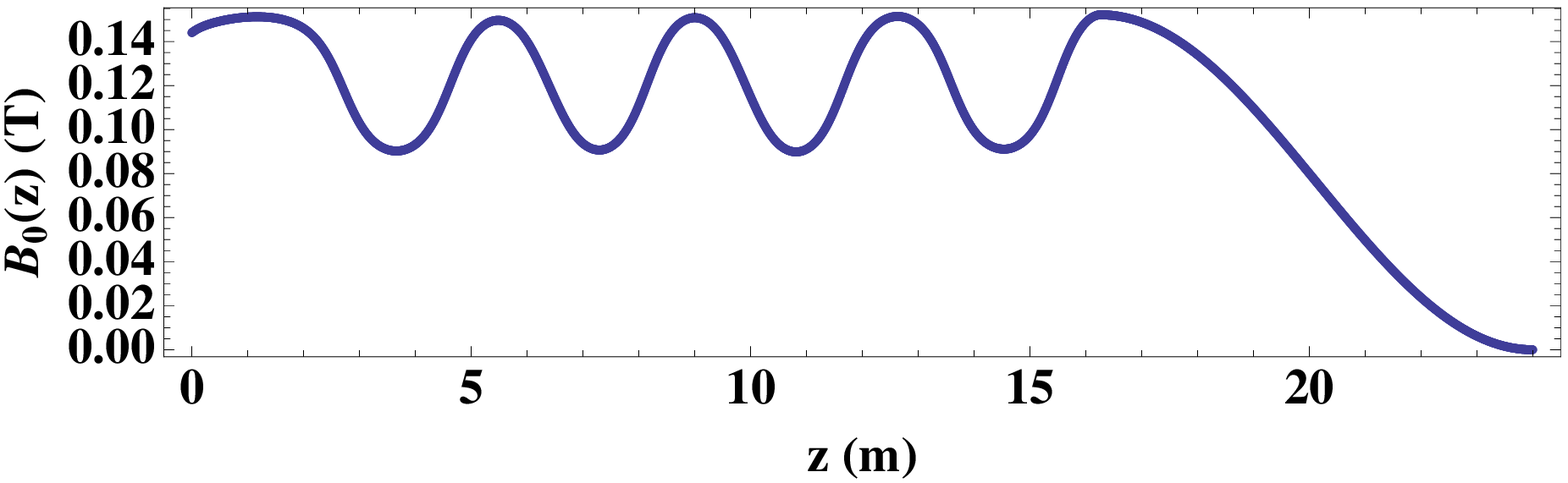}
\end{array}$
\end{center}
\caption{Computational domain (a) and static magnetic field configuration (b). The antenna is a blade antenna as used in the previous experiment to excite $m=0$ mode\cite{Zhang:2008aa}.}
\label{fg1}
\end{figure}

The computational domain and configuration of static magnetic field are shown in Fig.~\ref{fg1}, which are the same to those in \cite{Zhang:2008aa} except that now the coordinate is from left to right for convenient simulations. As employed previously, a magnetic beach is constructed as an absorbing end, and it is longer $7.74$~m (from $z=16.26$~m to $z=24$~m) than that in the experiment to ensure complete absorption; moreover, a finite ion collision frequency $\nu_i(z)$ has been introduced to resolve the ion cyclotron resonance in the beach area\cite{Zhang:2008aa}. Other parameters are taken from experiment directly, including helium ion specie, electron temperature $T_e=8$~eV, plasma density $n_e=9.2\times10^{17}\exp(-10.2~r^2)~\textrm{m}^{-3}$, blade antenna of length $L_a=1$~m and radius $R_a=0.005$~m, and diagnostic ports at $z=3.98$~m and $z=11.02$~m (or $10.25$~m and $3.21$~m respectively in the opposite coordinate). 

\begin{figure}[ht]
\begin{center}$
\begin{array}{ll}
(a)&(b)\\
\hspace{0.3cm}\includegraphics[width=0.48\textwidth,angle=0]{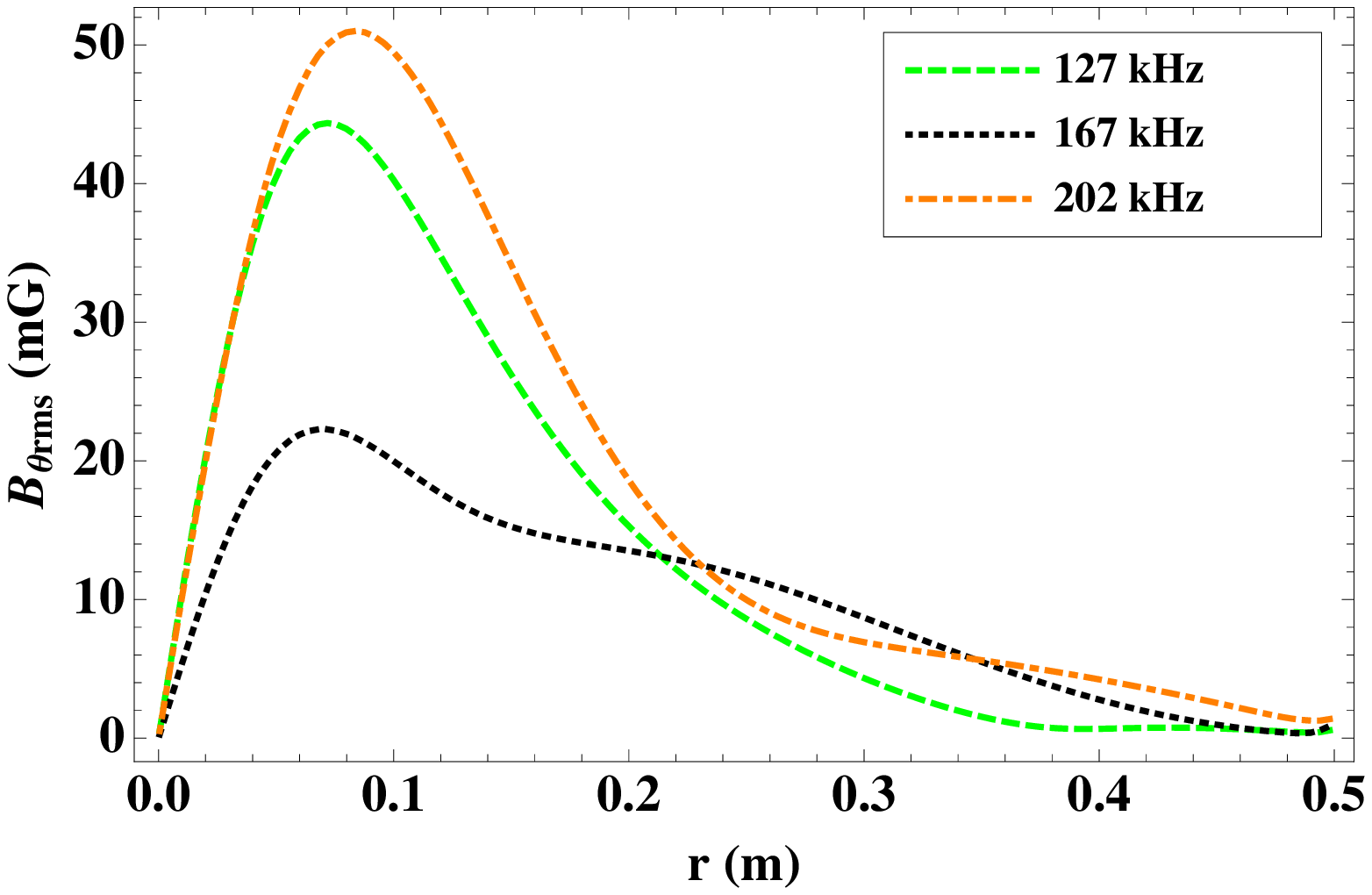}&\includegraphics[width=0.485\textwidth,angle=0]{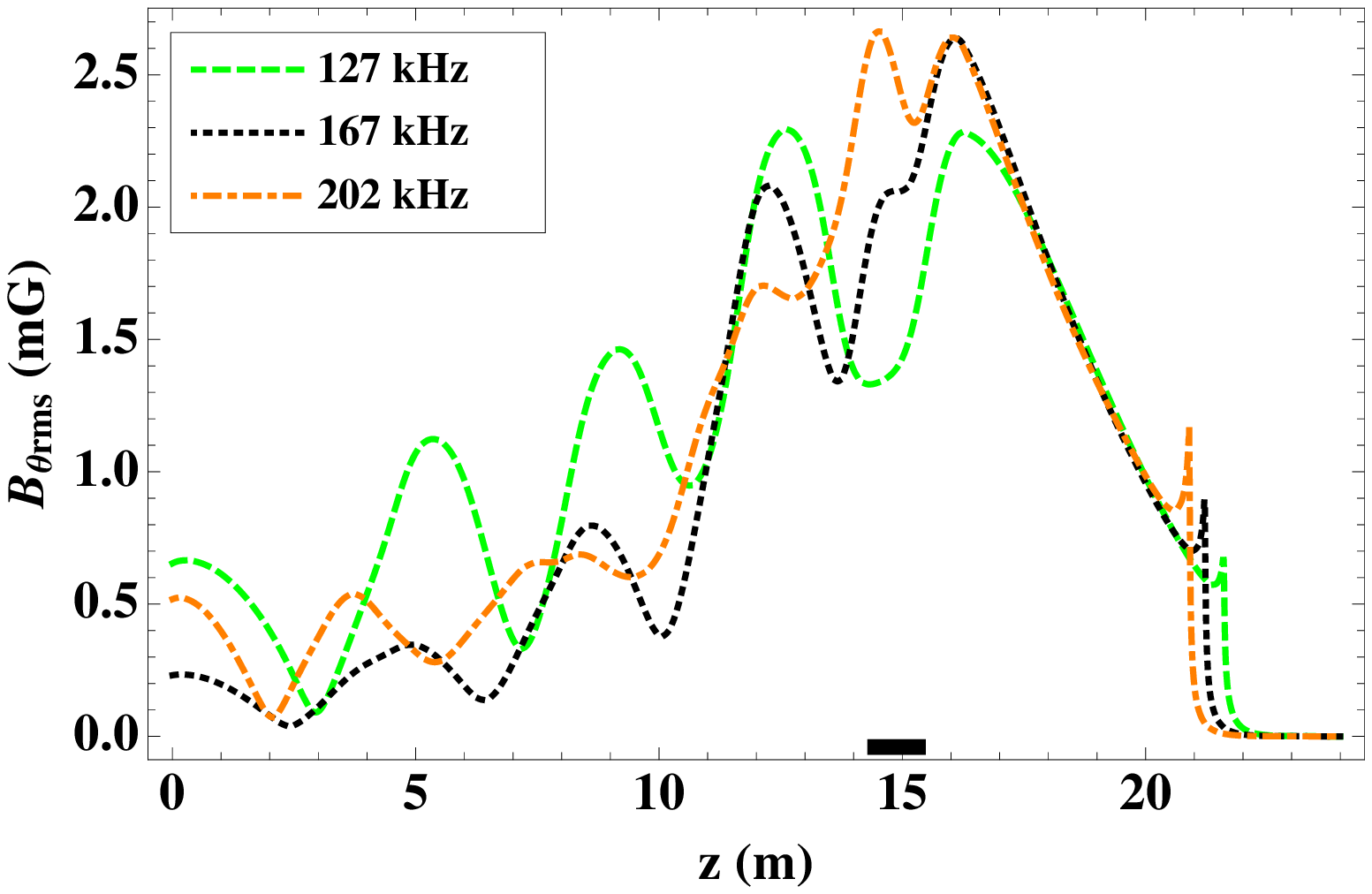}
\end{array}$
\end{center}
\caption{Variations of wave magnitude (rms) with radius ($z=3.98$~m) (a) and axis ($r=0$~m) (b) for three typical frequencies: $127$~kHz, $167$~kHz and $202$~kHz. The left figure is identical to Fig.~$14$ in the previous study\cite{Zhang:2008aa}. The black bar in the right figure labels the blade antenna.}
\label{fg2}
\end{figure}

\begin{figure}[ht]
\begin{center}$
\begin{array}{ll}
(a)&(b)\\
\hspace{-0.2cm}\includegraphics[width=0.49\textwidth,angle=0]{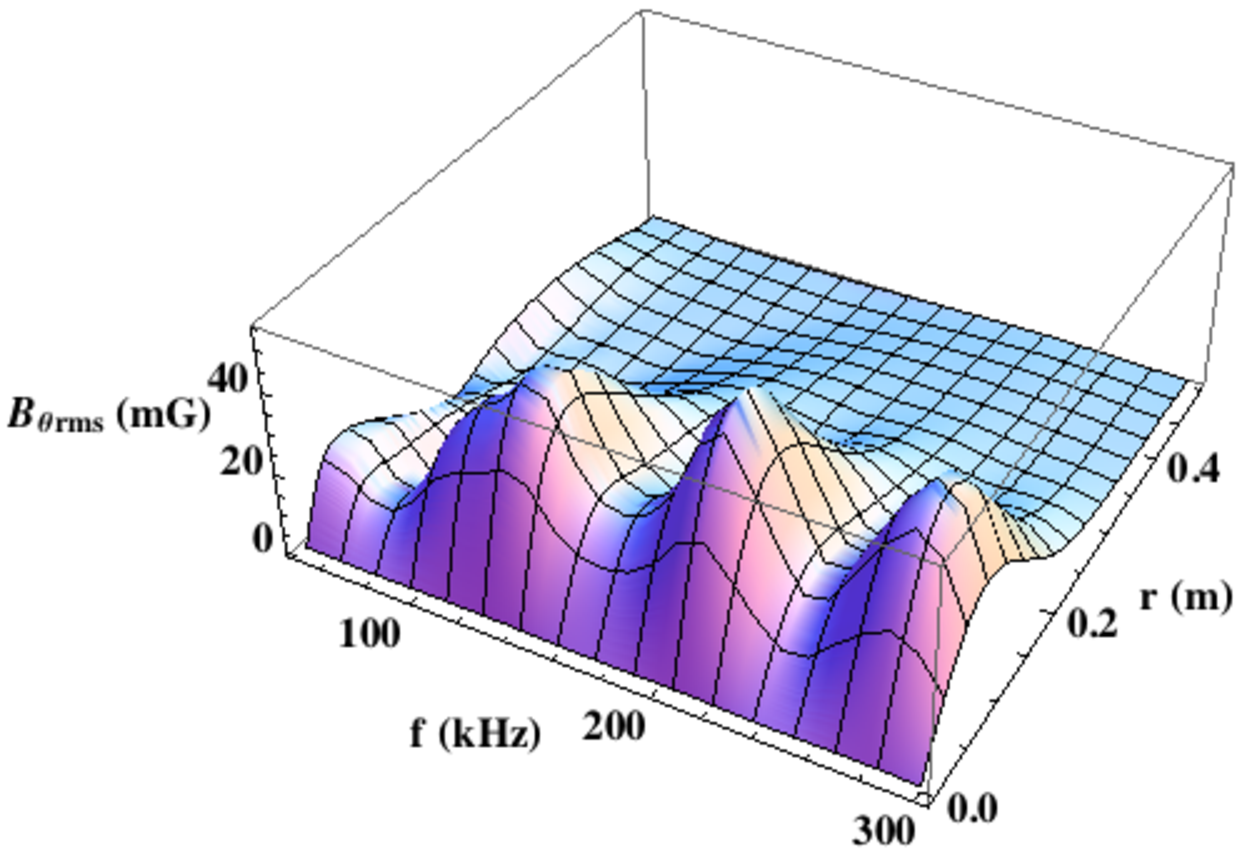}&\hspace{-0.1cm}\includegraphics[width=0.49\textwidth,angle=0]{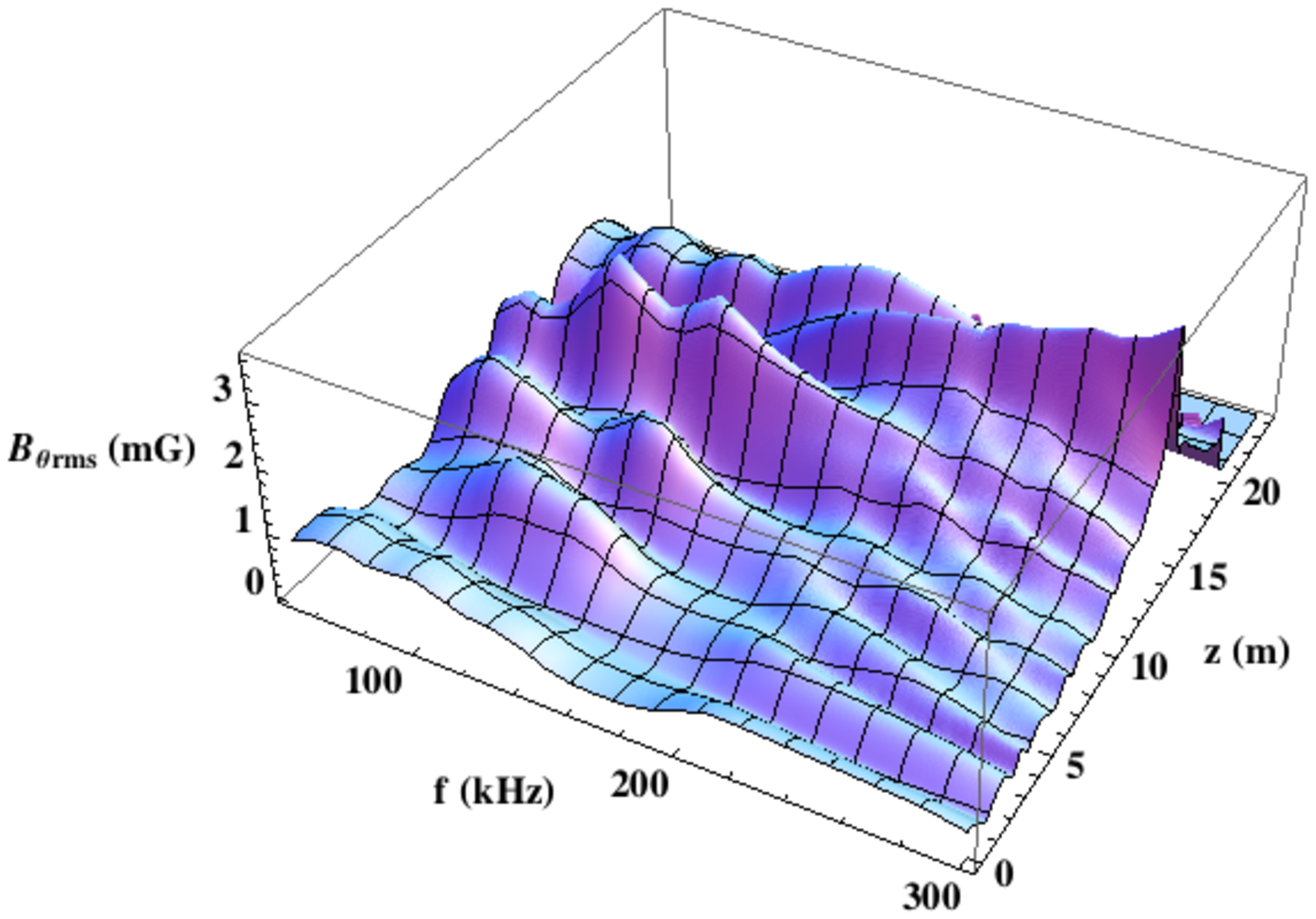}
\end{array}$
\end{center}
\caption{Surface plots of wave magnitude (rms) as a function of frequency and space location: (a) radius ($z=3.98$~m), (b) axis ($r=0$~m). The spectral gap is not globally clear as can be seen from (b).}
\label{fg3}
\end{figure}

\begin{figure}[ht]
\begin{center}$
\begin{array}{ll}
(a)&(b)\\
\hspace{0.3cm}\includegraphics[width=0.48\textwidth,angle=0]{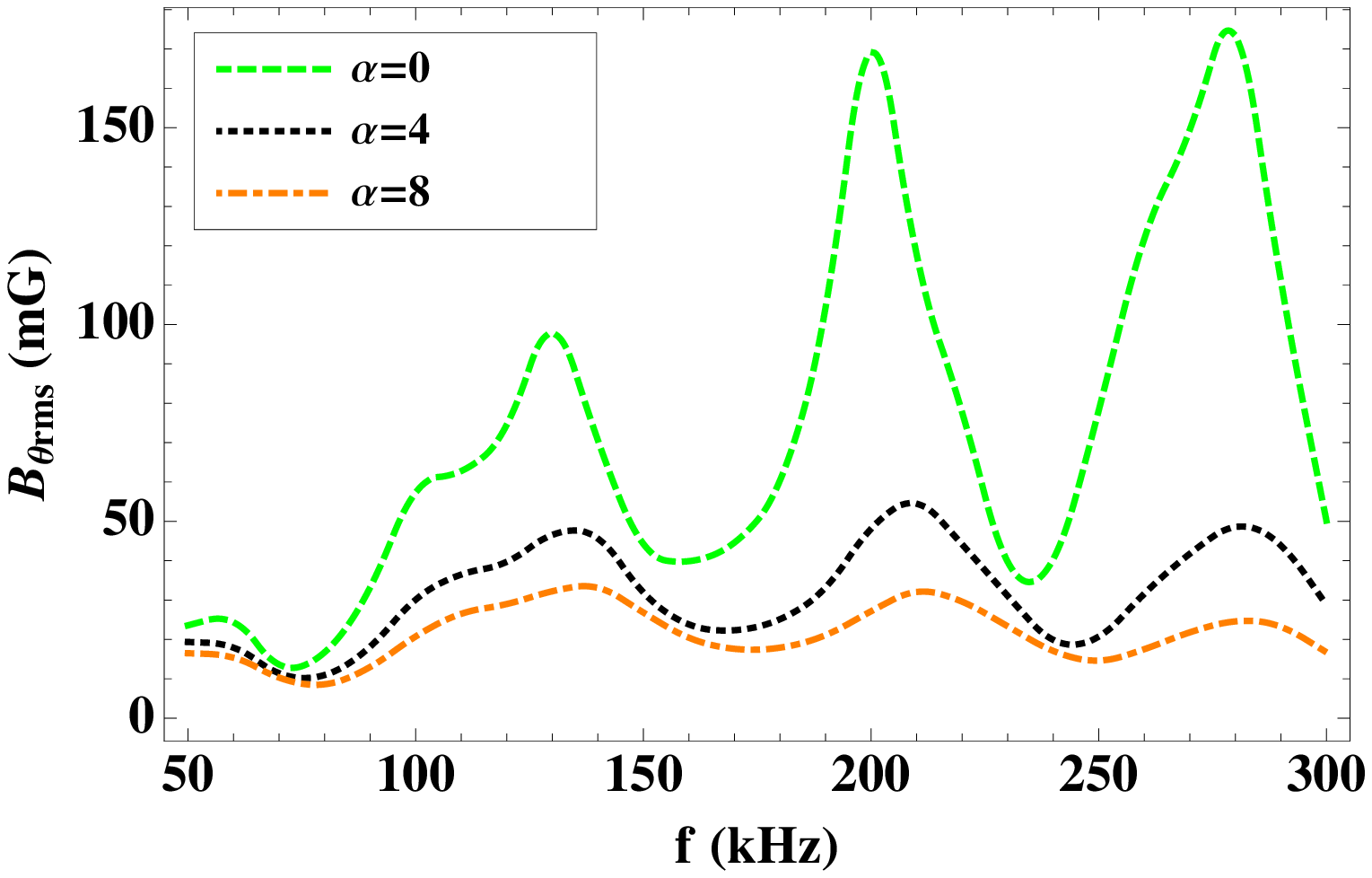}&\includegraphics[width=0.48\textwidth,angle=0]{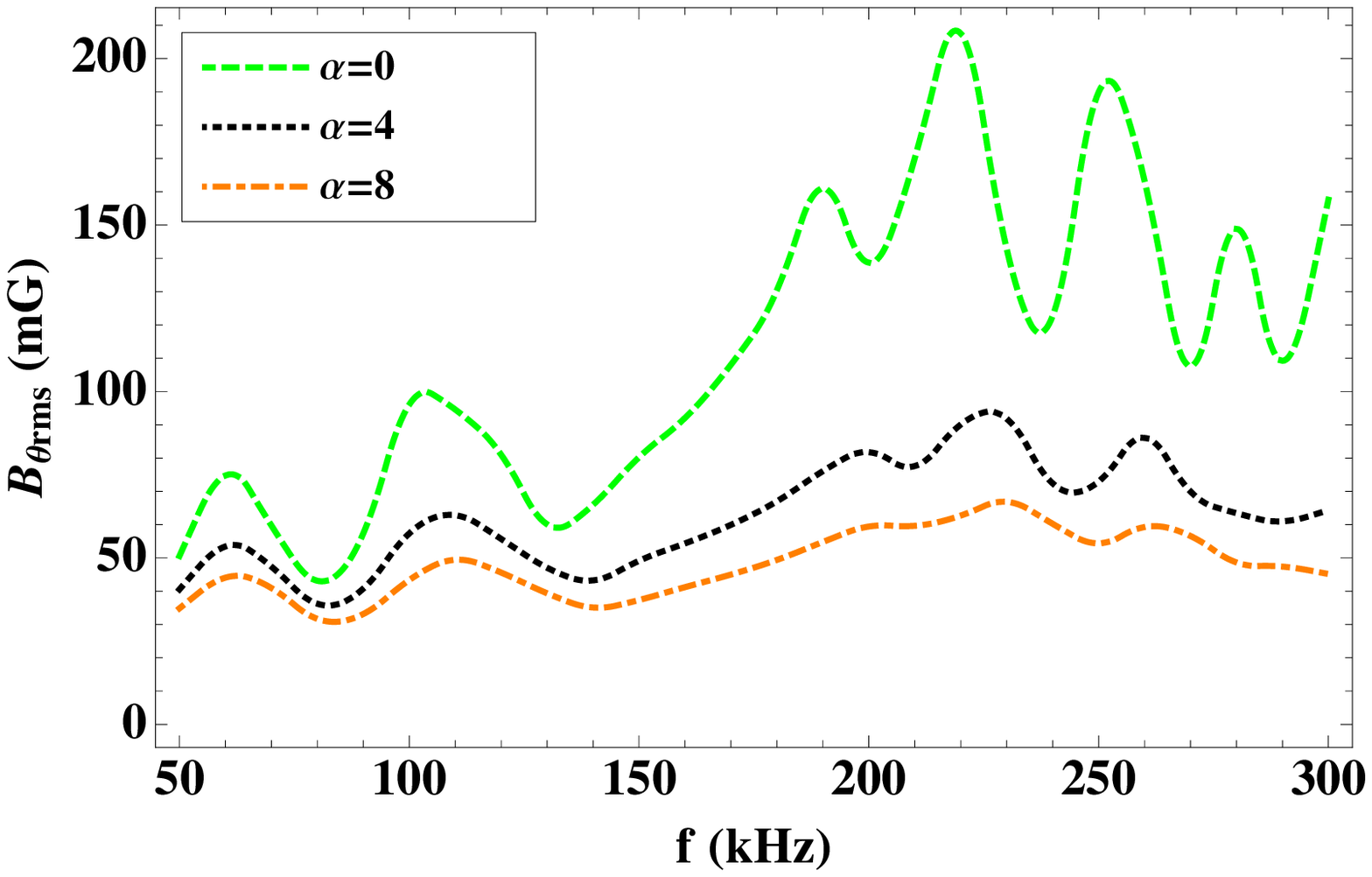}
\end{array}$
\end{center}
\caption{Variations of peak wave magnitude (rms) with frequency for three values of $\alpha$: $0$, $4$, $8$, which represent possible estimates of effective
collision frequencies for Landau damping, at $z=3.98$~m (a) and $z=11.02$~m (b). These two figures are identical to Fig.~$15$ in the previous study\cite{Zhang:2008aa}.}
\label{fg4}
\end{figure}

Figure~\ref{fg2} shows the variations of azimuthal wave magnetic field (rms) with radius and axis for three typical frequencies: $127$~kHz, $167$~kHz and $202$~kHz. The radial profiles and magnitudes of $B_{\theta \textrm{rms}}$ are identical to those previously published (Fig.~$14$ in \cite{Zhang:2008aa}), which exhibit a spectral gap centered around $167$~kHz and state the credibility of present simulations. The axial profiles of $B_{\theta \textrm{rms}}$ display a standing wave structure or beat pattern, implying strong reflections from the left endplate. Due to ion cyclotron resonance, the magnitude of $B_{\theta \textrm{rms}}$ decays quickly around $z=21$~m. 
A full picture of the wave field in the LAPD plasma is given by Fig.~\ref{fg3} in the spaces of $(f; r)$ and $(f; z)$. A strong and robust radial structure can be seen clearly from the $(f; r)$ space for different driving frequencies. However, a global spectral gap is not distinct in the $(f; z)$ space, which is surprising and indicates that the spectral gap observed in Fig.~\ref{fg2}(a) (also Fig.~\ref{fg5} below) may be an axially local result. 
This can be further confirmed by comparing the variations of peak wave magnitude (rms) with frequency at two axial locations, namely $z=3.98$~m and $z=11.02$~m, as shown in Fig.~\ref{fg4}. The values of $\alpha$ represent possible estimates of effective collision frequencies for Landau damping. The total effective collision frequency is termed as $\nu_{\textrm{eff}}=\nu_{ei}+\nu_{e\textrm{-Landau}}$ with $\nu_{e\textrm{-Landau}}=\alpha v_{\textrm{the}}\omega/v_A$\cite{Zhang:2008aa}. Except where acknowledged, $\alpha=4$ is used throughout the whole paper, and the Coulomb collision between ions and electrons $\nu_{ei}$ is believed to be the leading collision process affecting the energy deposition. The gap profiles measured at $z=3.98$~m are much better than those at $z=11.02$~m which exhibit nearly no spectral gap. Again, the curves in Fig.~\ref{fg4} are identical to those in Fig.~15 in \cite{Zhang:2008aa}, stating the credibility of our computations. 
Figure~\ref{fg5} displays the contour plots of wave energy density and power deposition density for three typical frequencies: $120$~kHz, $170$~kHz, $220$~kHz. Although Fig.~\ref{fg5}(a) shows a spectral gap centered around $170$~kHz, it is not thorough because waves can still propagate a distance (near the left endplate). This is also confirmed by the power deposition density shown in Fig.~\ref{fg5}(b), which is directly relevant to the wave field through energy conservation but does not show a clear spectral gap. The reason that this spectral gap is not thorough and global may be due to very few magnetic mirrors and small modulation depth employed in the previous experiment. Moreover, the experimentally constructed magnetic mirror array shown in Fig.~\ref{fg1}(b) is not perfectly periodic. The axial length of each mirror is not exactly $3.63$~m but with $3-18.7\%$ difference, and the maxima and minima also have about $1\%$ difference around $0.15$~T and $0.09$~T respectively. Although these differences are small, they can effect the dispersion relation of Alfv\'{e}n waves significantly. 
\begin{figure}[ht]
\begin{center}$
\begin{array}{ll}
(a)&(b)\\
\includegraphics[width=0.48\textwidth,angle=0]{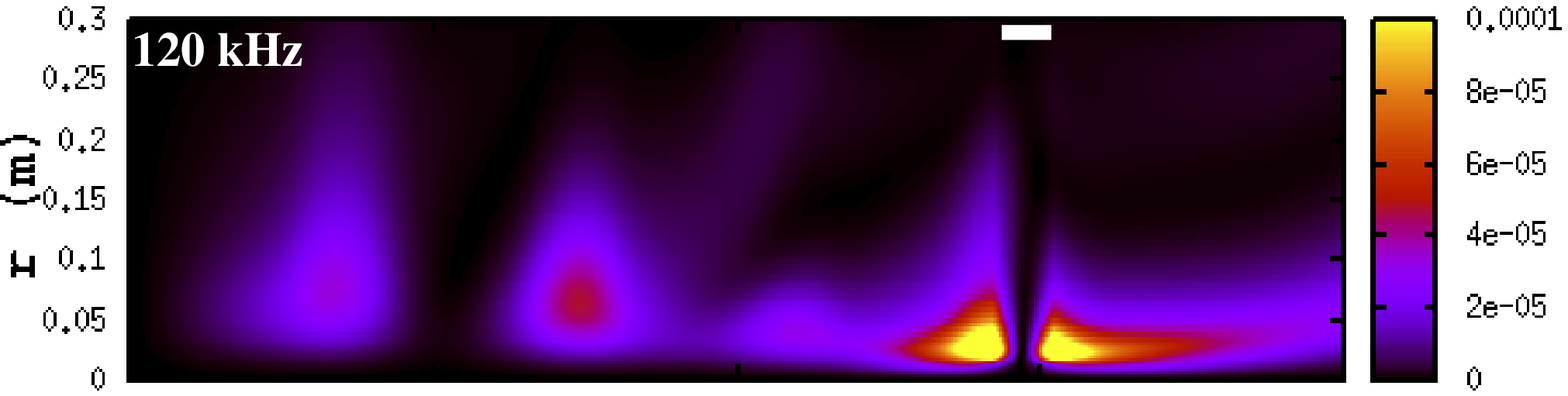}&\includegraphics[width=0.48\textwidth,angle=0]{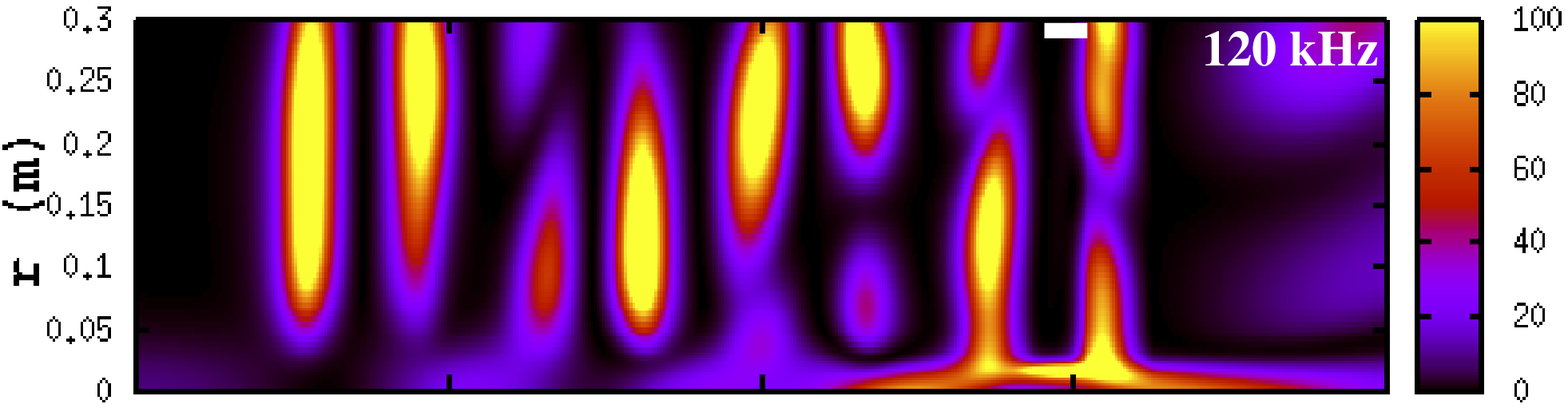}\\
\includegraphics[width=0.48\textwidth,angle=0]{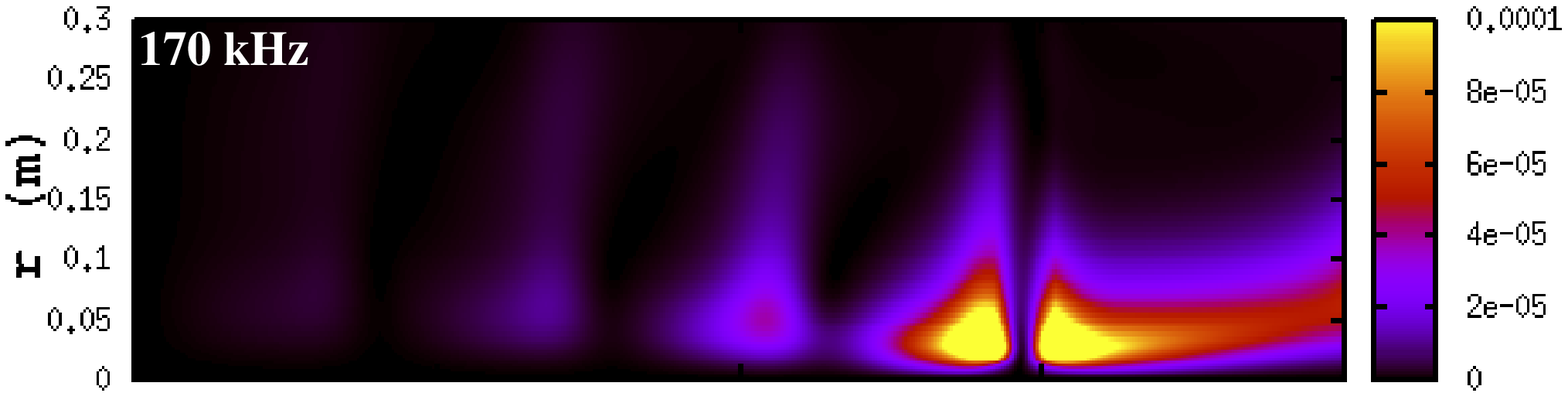}&\includegraphics[width=0.48\textwidth,angle=0]{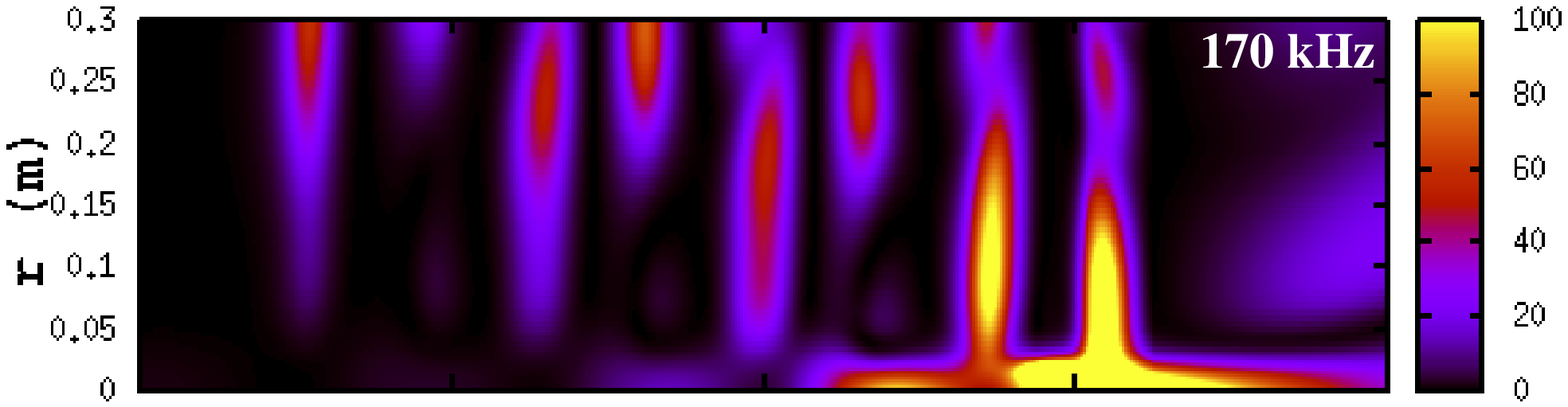}\\
\includegraphics[width=0.48\textwidth,angle=0]{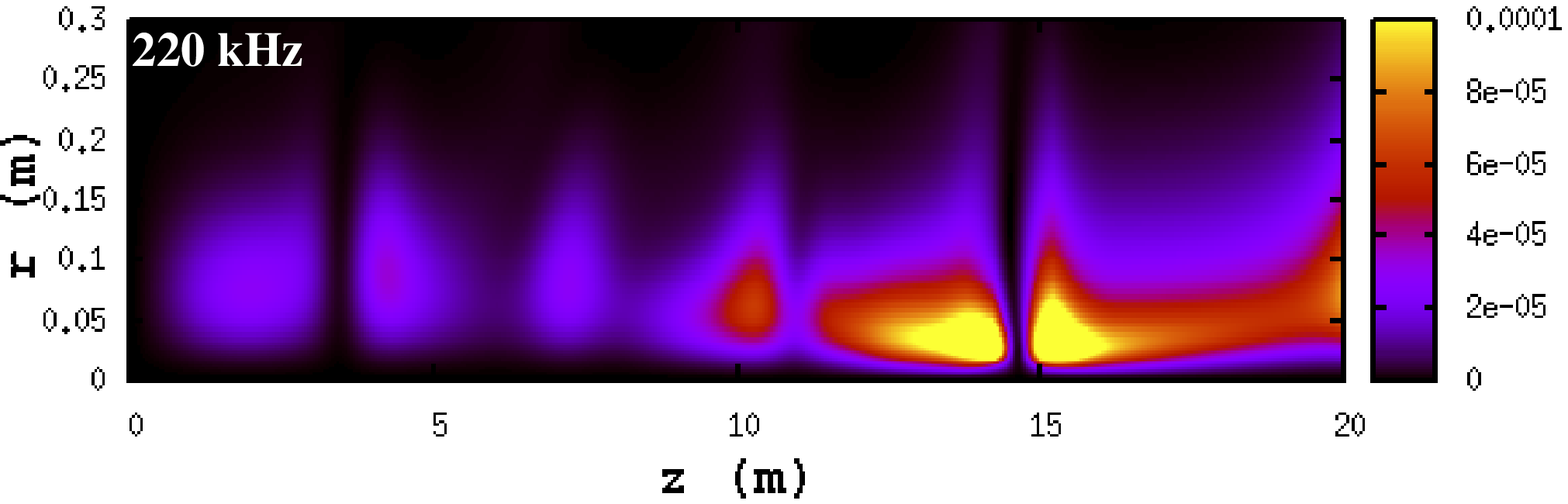}&\includegraphics[width=0.48\textwidth,angle=0]{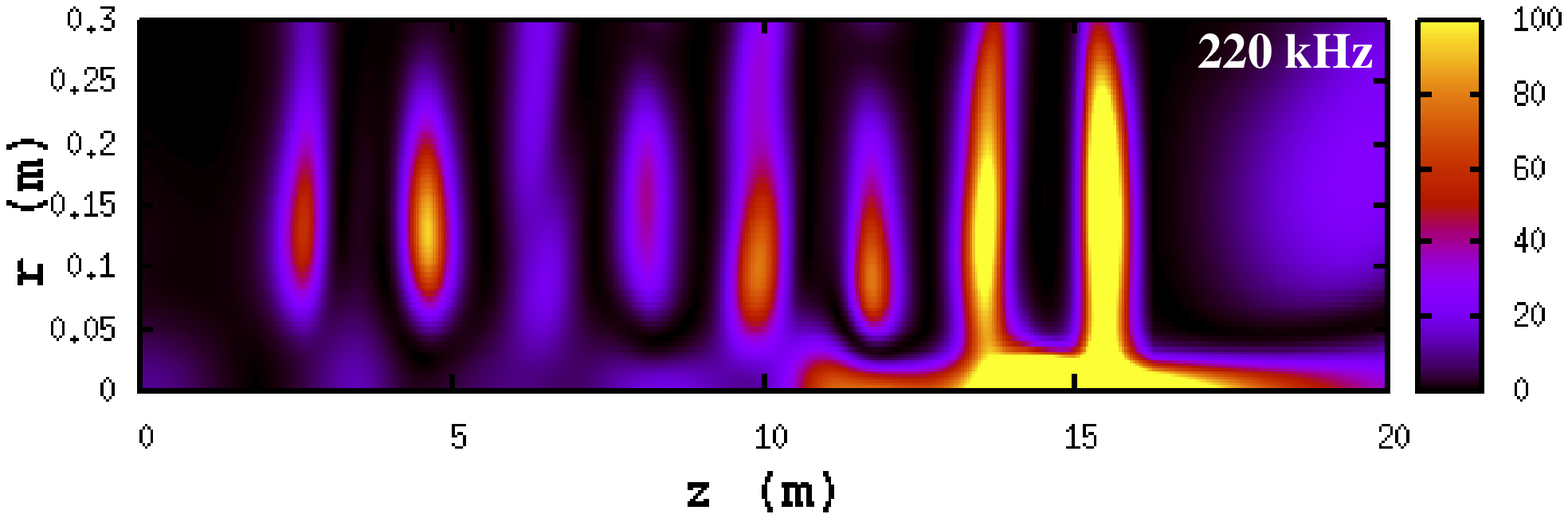}
\end{array}$
\end{center}
\caption{Contour plots of wave energy density ($J/m^3$) (a) and power deposition density ($W/m^3$) (b) for three typical frequencies: $120$~kHz, $170$~kHz, $220$~kHz. The left figure is identical to Fig.~$16$ in the previous study\cite{Zhang:2008aa}. The white bar labels the axial location of antenna.}
\label{fg5}
\end{figure}
Figure~\ref{fg6} shows the dispersion curves for the experimental magnetic mirror shown in Fig.~\ref{fg1}(b) and numerical magnetic mirror of $B_0(z)=0.12+0.3\cos[2\pi(z-16.84)/3.63]$, together with the analytical dispersion relation of $\omega/k_z=v_A=B_0/\sqrt{\mu_0 m_i n_{i0}}$ for uniform magnetic field (here $n_{i0}$ is the plasma density on axis). It can be seen that the imperfect periodicity shifts the dispersion curve to lower frequency range and distorts the curve shape. 
\begin{figure}[ht]
\begin{center}
\includegraphics[width=0.55\textwidth,angle=0]{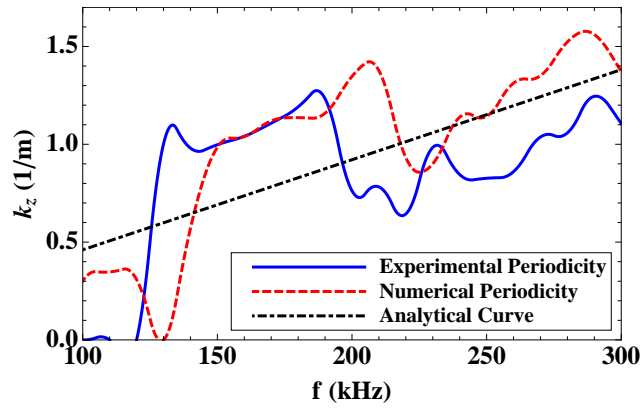}
\end{center}
\caption{Dispersion relations for experimentally imperfect periodicity and numerically perfect periodicity, together with the analytical dispersion relation of $\omega/k_z=v_A=B_0/\sqrt{\mu_0 m_i n_{i0}}$ in a slab geometry.}
\label{fg6}
\end{figure}

\section{Global spectral gap and gap eigenmode}\label{age}
\subsection{High field strength and plasma density}
\begin{figure}[ht]
\begin{center}
\includegraphics[width=0.735\textwidth,angle=0]{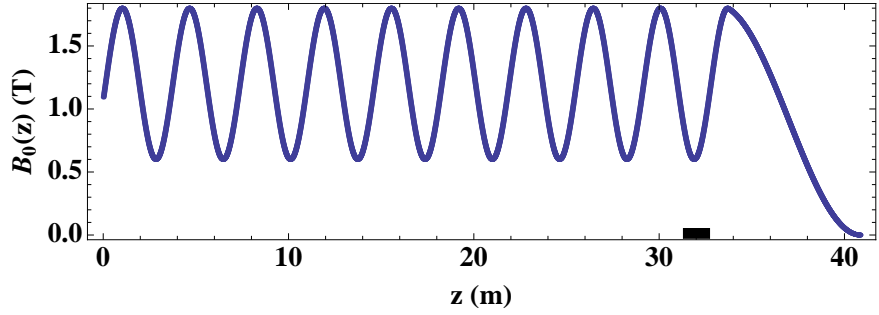}
\end{center}
\caption{Configuration of numerically constructed magnetic field with periodic part of $B_0(z)=1.2+0.6\cos[2\pi(z-33.68)/3.63]$ and magnetic beach part of $7.74$~m. The black bar labels the location of the blade antenna.}
\label{fg7}
\end{figure}

\begin{figure}[ht]
\begin{center}$
\begin{array}{ll}
(a)&(b)\\
\hspace{-0.2cm}\includegraphics[width=0.49\textwidth,angle=0]{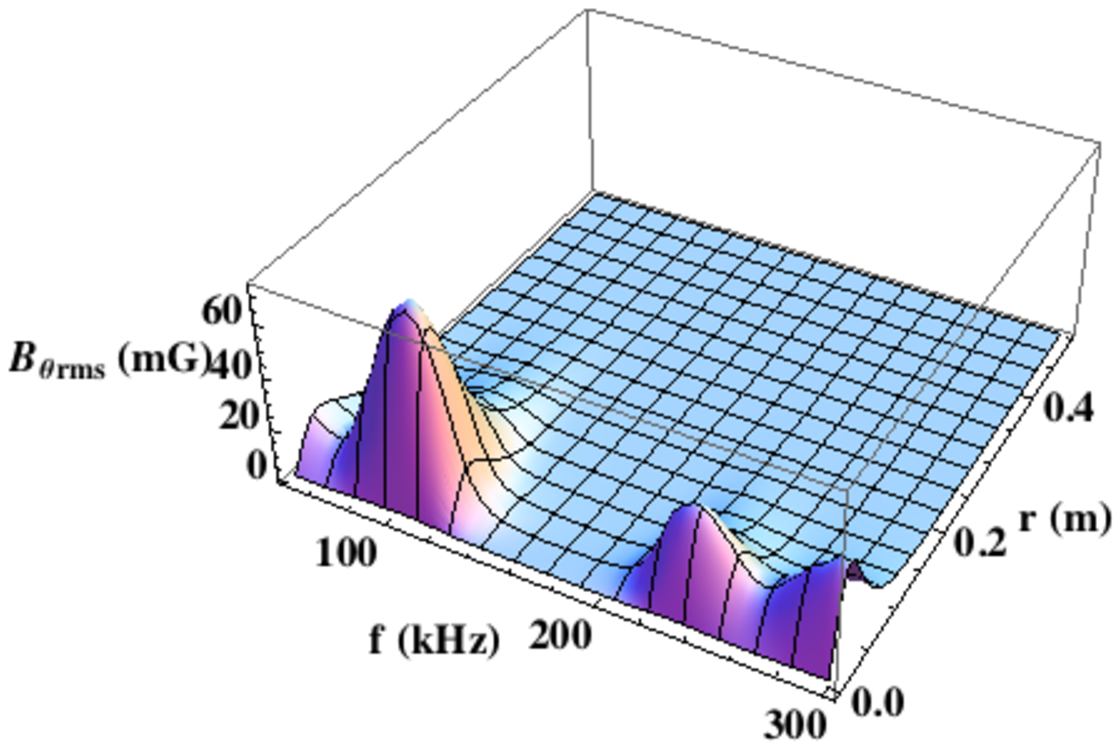}&\hspace{-0.1cm}\includegraphics[width=0.49\textwidth,angle=0]{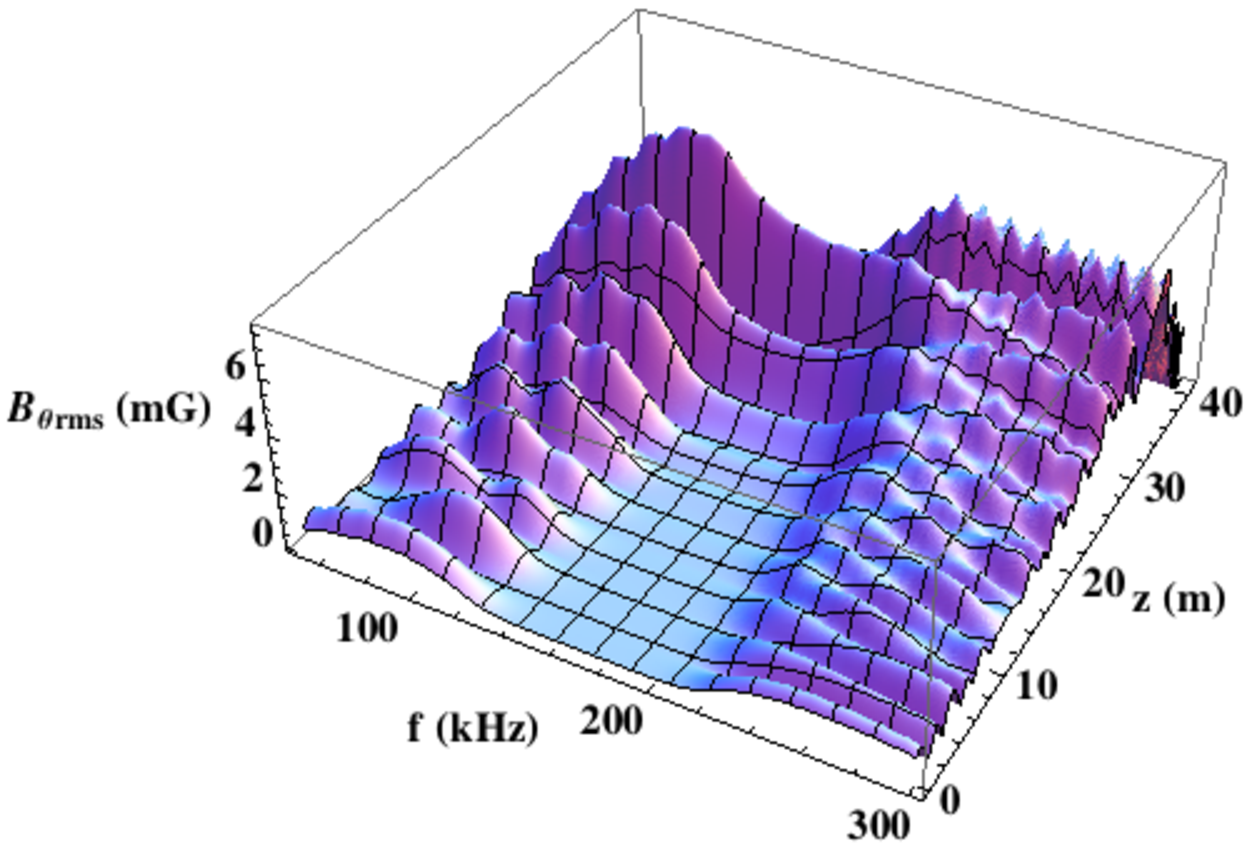}
\end{array}$
\end{center}
\caption{Surface plots of wave magnitude (rms) as a function of frequency and space location: (a) radius ($z=3.98$~m), (b) axis ($r=0$~m). A global spectral gap can be seen clearly both in radial and axial directions.}
\label{fg8}
\end{figure}

Based on previous analyses, this section will increase the number of magnetic mirrors, which can be done only by increasing the length of mirror area due to the unchangeable minimum length of each mirror, namely $3.63$~m on the LAPD, and construct the mirror field numerically with perfect periodicity. To broaden the width of spectral gap for easy gap eigenmode formation, the modulation depth will be increased as well according to $\Delta\omega=\epsilon\omega_0$ in previous studies with $\epsilon$ the modulation depth\cite{Zhang:2008aa, Chang:2013aa, Chang:2014aa}. Furthermore, because the collisional damping rate of Alfv\'{e}n waves can be approximated as $3\omega^2\nu_{ei}/(2\omega_{ci}|\omega_{ce}|)$\cite{Braginskii:1965aa, Chang:2014aa}, the magnitude of field strength will be increased to have large ion cyclotron frequency and thereby low Alfv\'{e}nic damping rate. The final constructed periodic magnetic field is $B_0(z)=1.2+0.6\cos[2\pi(z-33.68)/3.63]$. The length of mirror area and modulation depth are doubled, and the field strength is increased by $10$ times. Here, to maintain the phase velocity of Alfv\'{e}n waves for comparison with previous work, the plasma density is increased by $100$ times ($v_A=B_0/\sqrt{\mu_0 m_i n_{i0}}$). The numerically constructed magnetic field with absorbing beach is shown in Fig.~\ref{fg7}. 
Figure~\ref{fg8} gives the surface plots of computed wave field in the spaces of $(f; r)$ and $(f; z)$, both of which show a global spectral gap. 
The center and width of this spectral gap largely agree with a simple analytical estimate from $\omega/k_z=v_A=B_0/\sqrt{\mu_0 m_i n_{i0}}$ and $\omega_\pm=\omega_0(1\pm\epsilon/2)$\cite{Zhang:2008aa, Chang:2013aa, Chang:2014aa}, which gives $f=188\pm47$~kHz. 

\begin{figure}[ht]
\begin{center}$
\begin{array}{ll}
(a)&(b)\\
\hspace{-0.2cm}\includegraphics[width=0.49\textwidth,angle=0]{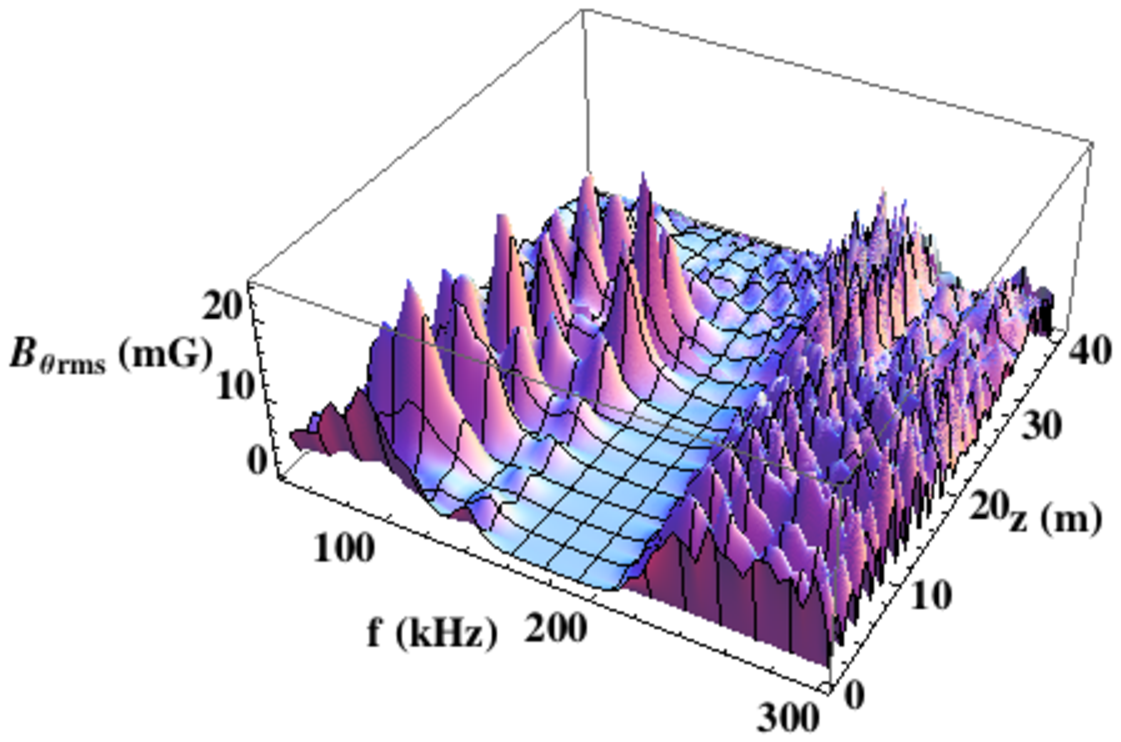}&\hspace{-0.1cm}\includegraphics[width=0.49\textwidth,angle=0]{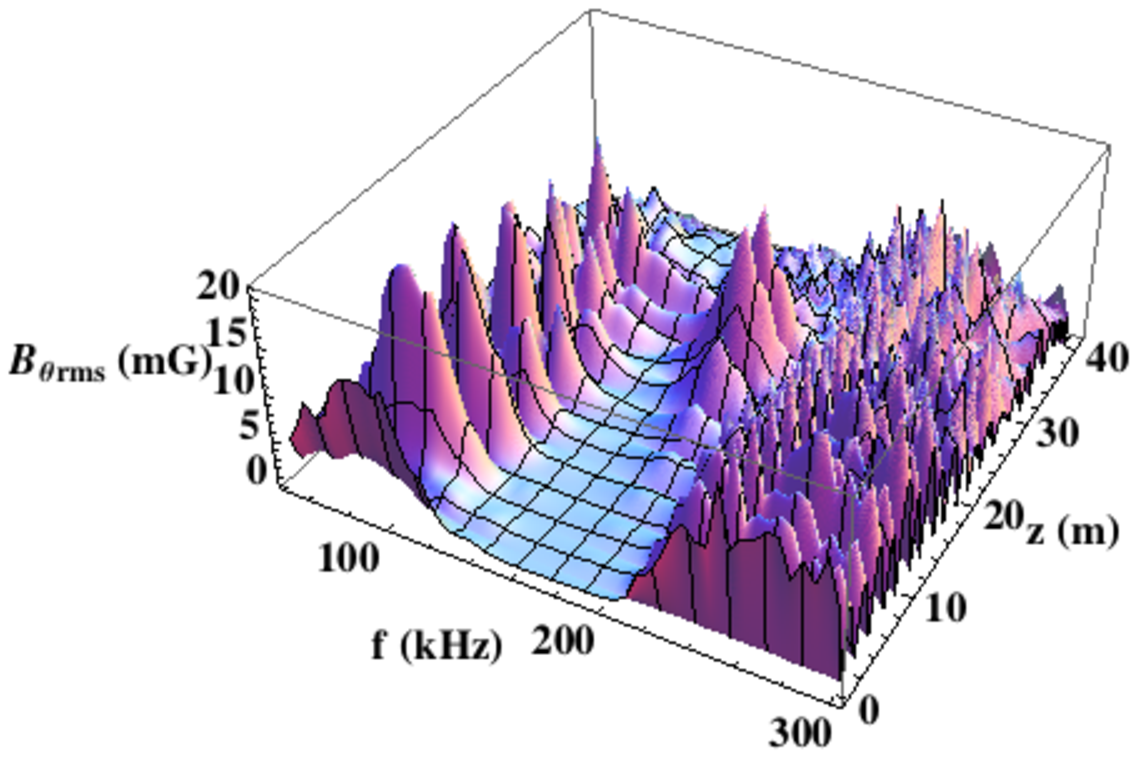}
\end{array}$
\end{center}
\caption{Surface plots of wave magnitude (rms) as a function of frequency and axial location ($r=0$~m): (a) odd-parity, (b) even-parity. Gap eigenmodes are clearly formed inside the spectral gap.}
\label{fg10}
\end{figure}

\begin{figure}[ht]
\begin{center}$
\begin{array}{ll}
(a_1)&(b_1)\\
\hspace{0.35cm}\includegraphics[width=0.47\textwidth,angle=0]{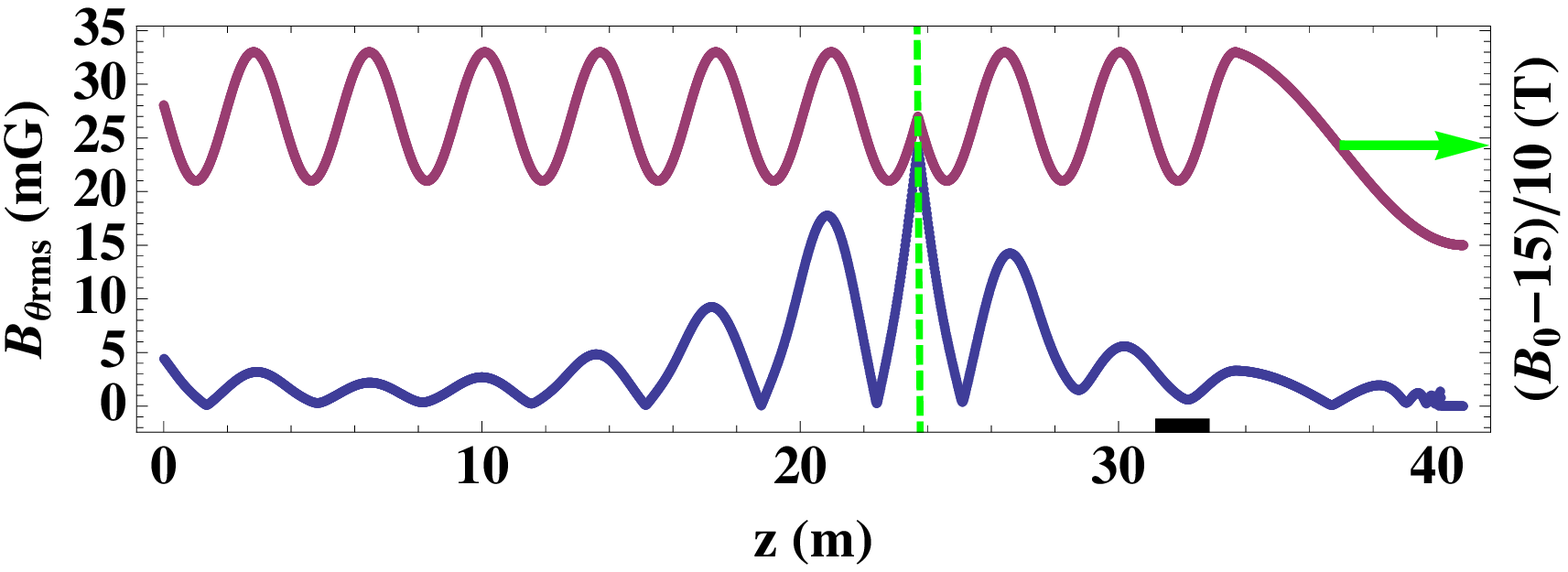}&\hspace{-0.25cm}\includegraphics[width=0.478\textwidth,angle=0]{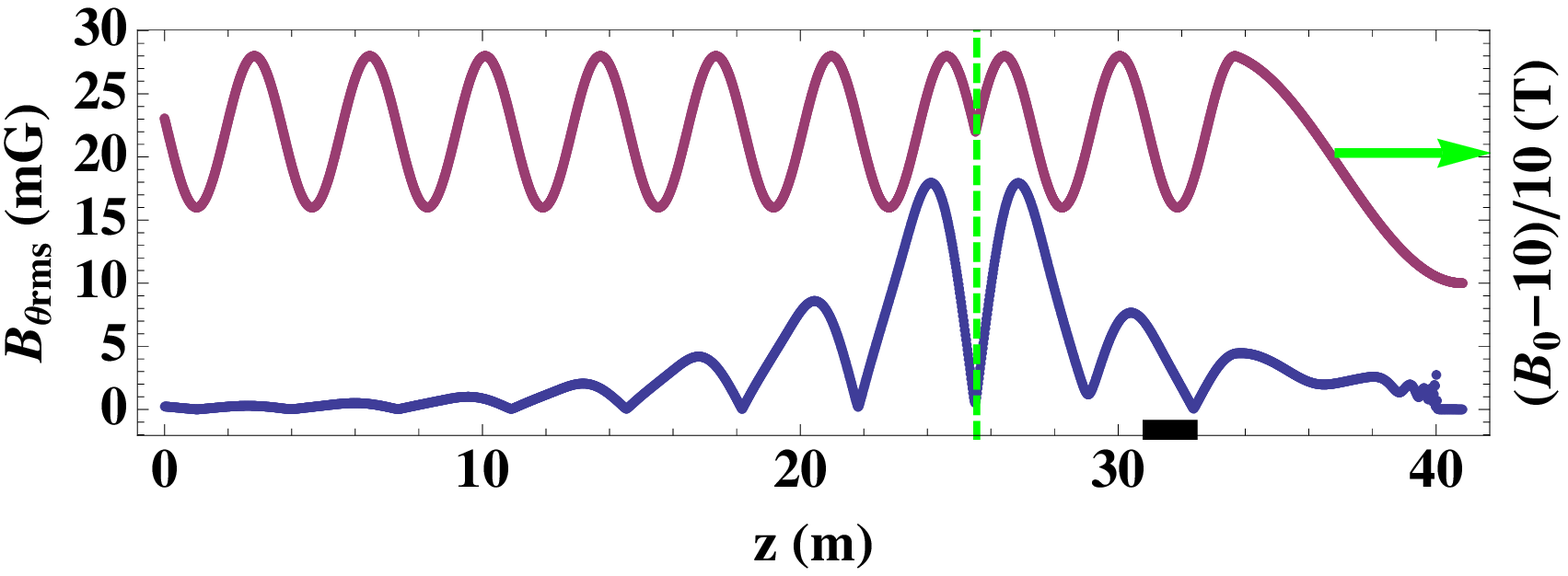}\\
(a_2)&(b_2)\\
\includegraphics[width=0.53\textwidth,angle=0]{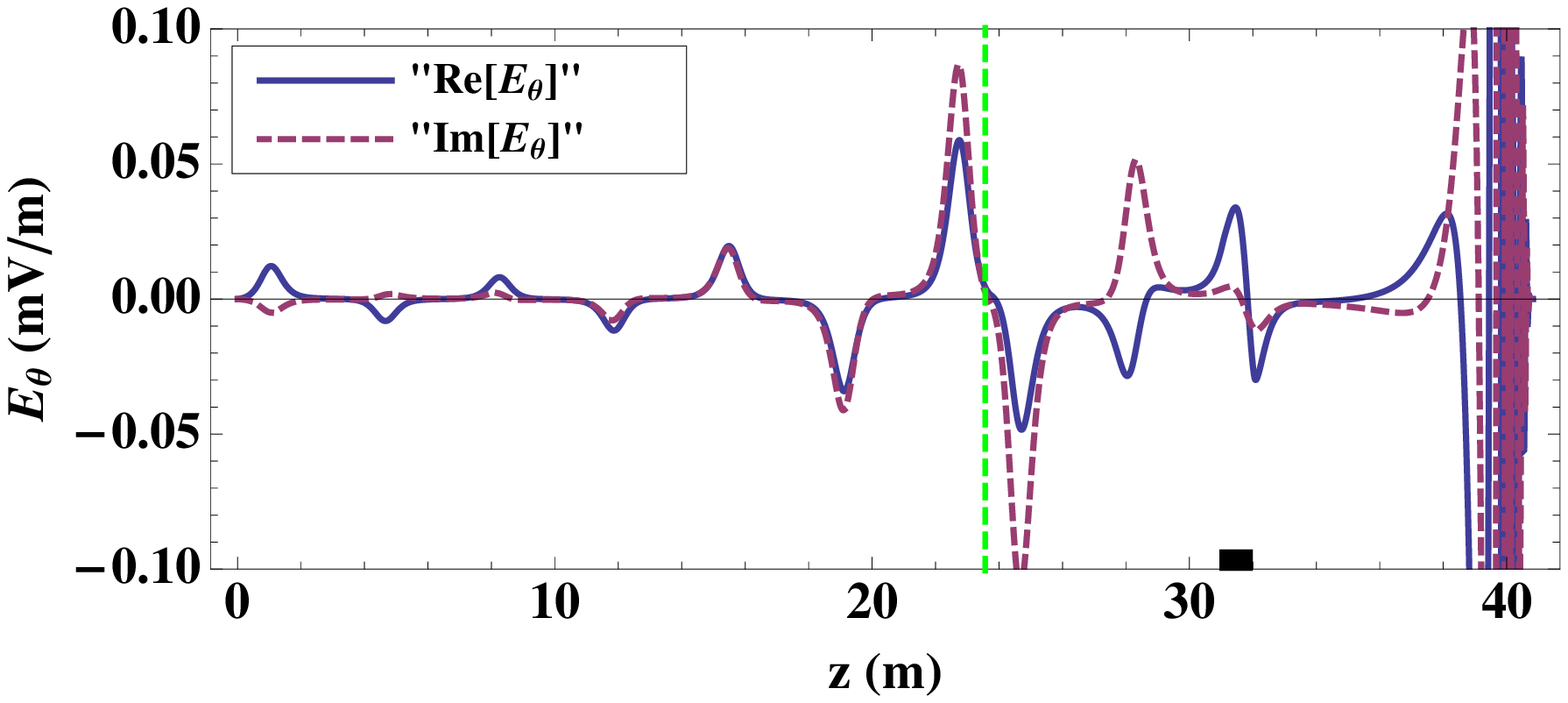}&\hspace{-0.6cm}\includegraphics[width=0.535\textwidth,angle=0]{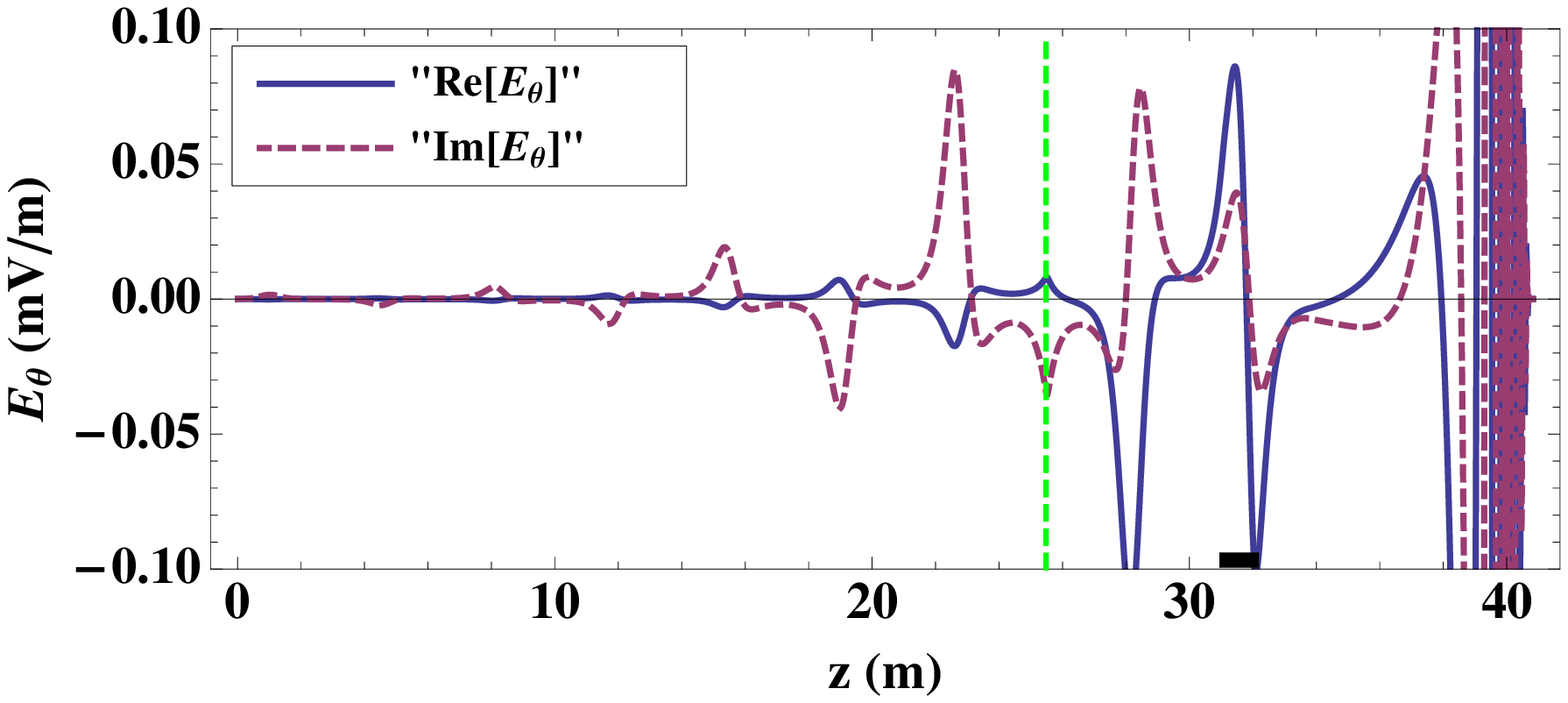}
\end{array}$
\end{center}
\caption{Axial profiles ($r=0$~m) of the azimuthal components of wave magnetic field ($a_1$ and $b_1$) and wave electric field ($a_2$ and $b_2$) for odd-parity ($140$~kHz) and even-parity ($185$~kHz) peak gap eigenmodes. The black bar and green dot-dashed line label the locations of external driving antenna and periodicity-broken defect, respectively. The odd-parity and even-parity phrases characterize the wave electric field across these defects at $z=23.6975$~m and $z=25.5125$~m, respecrtively.}
\label{fg11}
\end{figure}

Following a similar strategy as used before\cite{Chang:2013aa, Chang:2014aa}, two types of gap eigenmodes can be formed inside the spectral gap, namely odd-parity and even-parity for two types of defects, as shown by the surface plots in Fig.~\ref{fg10}. Here the collisionality has been decreased by a factor of $10$ to obtain sharp resoance peaks. 
The employed periodic static magnetic field with local defects are shown in Fig.~\ref{fg11}($a_1$) and Fig.~\ref{fg11}($b_1$), together with the axial profiles of peak gap eigenmodes. The parity is featured by the axial profile of $E_\theta$ across the defect location as illustrated in Fig.~\ref{fg11}($a_2$) and Fig.~\ref{fg11}($b_2$). It can be seen that the wave length of formed gap eigenmode is nearly twice the periodicity of external magnetic mirror, which is consistent with the Bragg's law. Moreover, the eigenmode is a standing wave localized around the defect, same as observed previously\cite{Chang:2013aa, Chang:2014aa}. The odd-parity and even-parity gap eigenmodes peak at $f=140$~kHz and $f=185$~kHz, respectively, in the range of $50-300$~kHz with resolution of $5$~kHz. 

To implement these gap eigenmodes on LAPD, there are three limitations considering the present experimental capability: field strength, plasma density and machine length. The maximum field strength is $0.3$~T at present\cite{Gekelman:1991aa}, thus it may need super conducting magnets to become $6$ times stronger, namely $1.8$~T.\cite{Squire:2007aa} More input power or innovative ionization method is also required to enhance the plasma density from maximum $2\times10^{19}~\textrm{m}^{-3}$ now to $9.2\times10^{19}~\textrm{m}^{-3}$.\cite{Squire:2007aa, Xia:2017aa} The length of mirror area could be doubled by setting up a reflecting endplate which turns back Alfv\'{e}n waves to propagate again through magnetic mirror array. An alternative way is to reduce the periodicity of each mirror to be half, namely $3.63/2$~m. The odd-parity and even-parity mode features correspond to boundary conditions of $E_\theta(z_0)=0$ and $E'_\theta(z_0)=0$ respectively\cite{Chang:2013aa, Chang:2014aa}, and can be implemented by a conducting mesh and conducting ring in experiment, whose radial and axial slots are then required accordingly to prevent the formation of azimuthal current\cite{Chang:2014aa}. Hence, these limitations are solvable in experiment. The following section will be devoted to further removing these limitations by reducing the field strength, plasma density and the number of magnetic mirrors. 

\subsection{Low field strength and plasma density}
\begin{figure}[ht]
\begin{center}$
\begin{array}{ll}
(a_1)&(b_1)\\
\includegraphics[width=0.49\textwidth,angle=0]{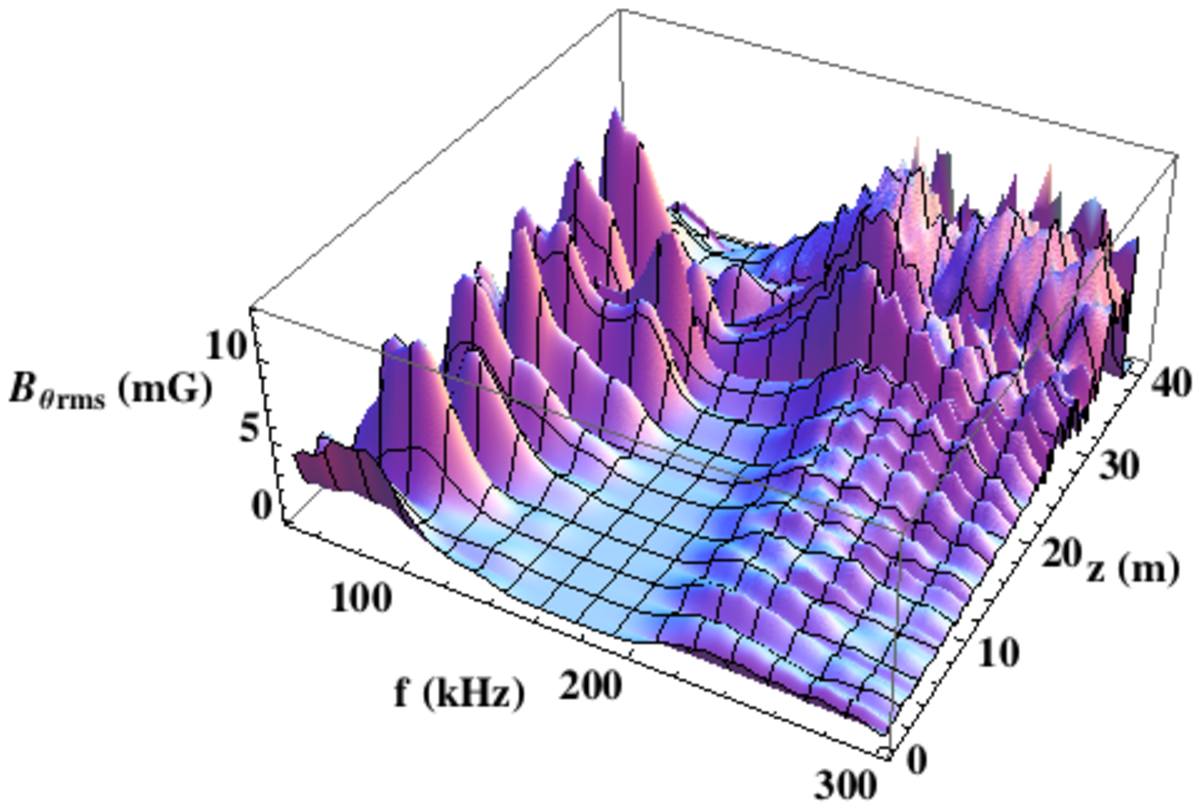}&\includegraphics[width=0.49\textwidth,angle=0]{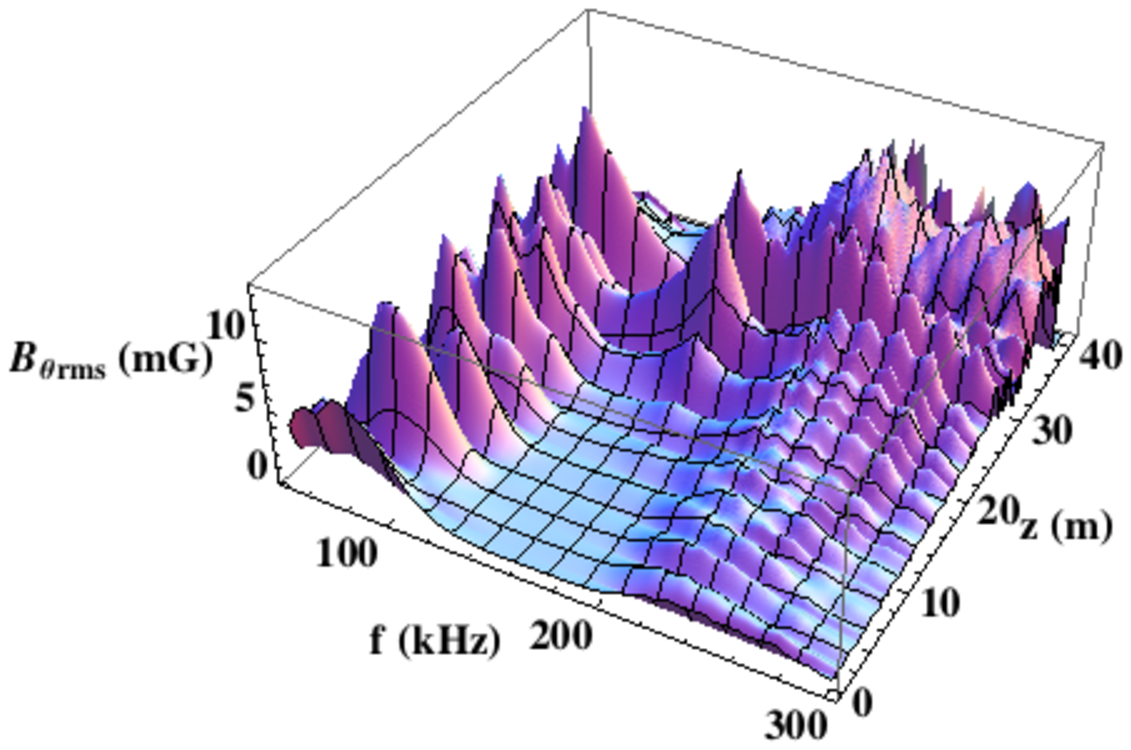}\\
(a_2)&(b_2)\\
\includegraphics[width=0.49\textwidth,angle=0]{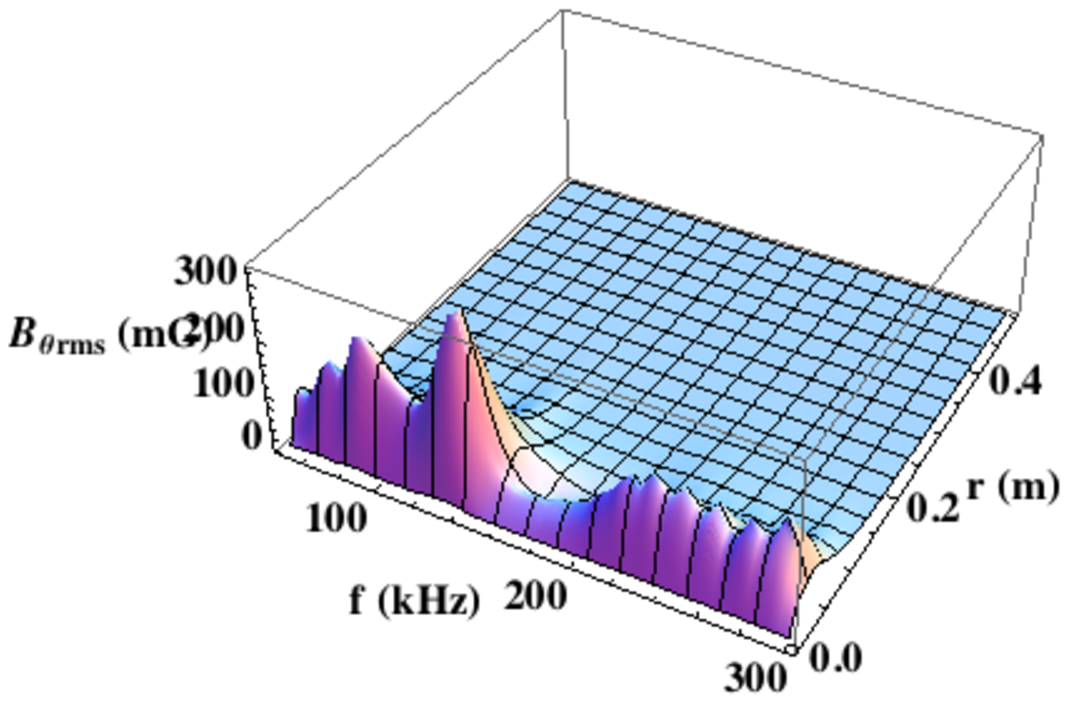}&\includegraphics[width=0.49\textwidth,angle=0]{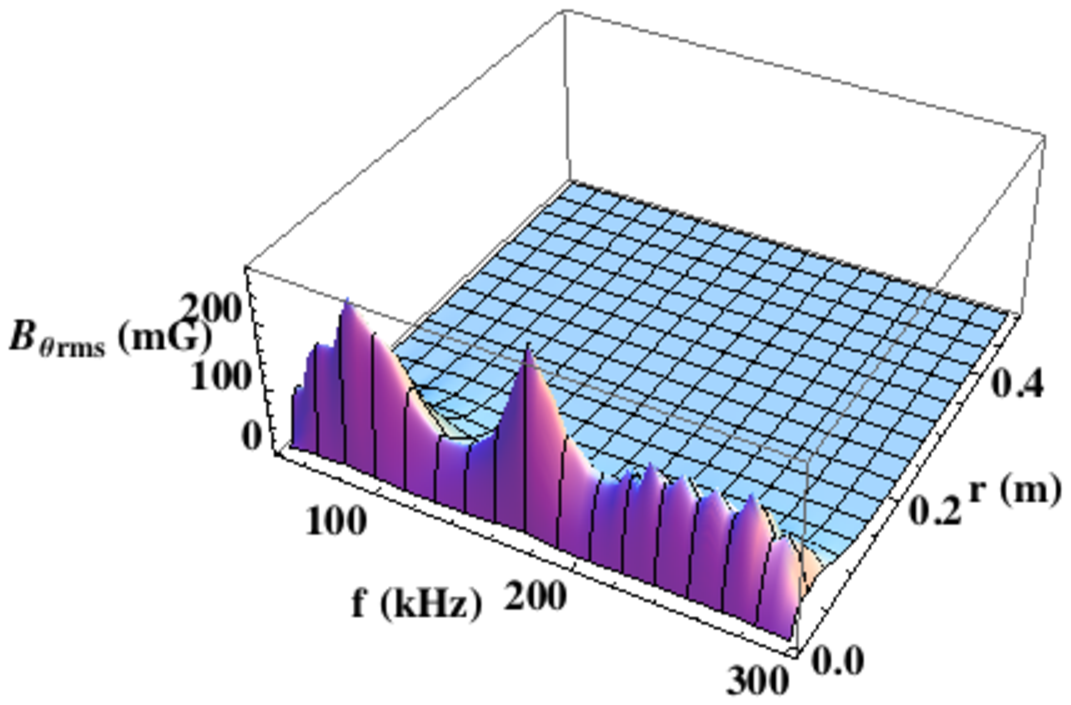}
\end{array}$
\end{center}
\caption{Surface plots of wave magnitude (rms) as a function of frequency and axial location ($a_1$ and $b_1$ at $r=0$~m) or radial location ($a_2$ at $z=23.6975$~m and $b_2$ at $z=25.5125$~m). The left figures (a) are odd-parity gap eigenmode while the right right figures (b) are even-parity gap eigenmode.}
\label{fg12}
\end{figure}

\begin{figure}[ht]
\begin{center}$
\begin{array}{ll}
(a)&(b)\\
\includegraphics[width=0.49\textwidth,angle=0]{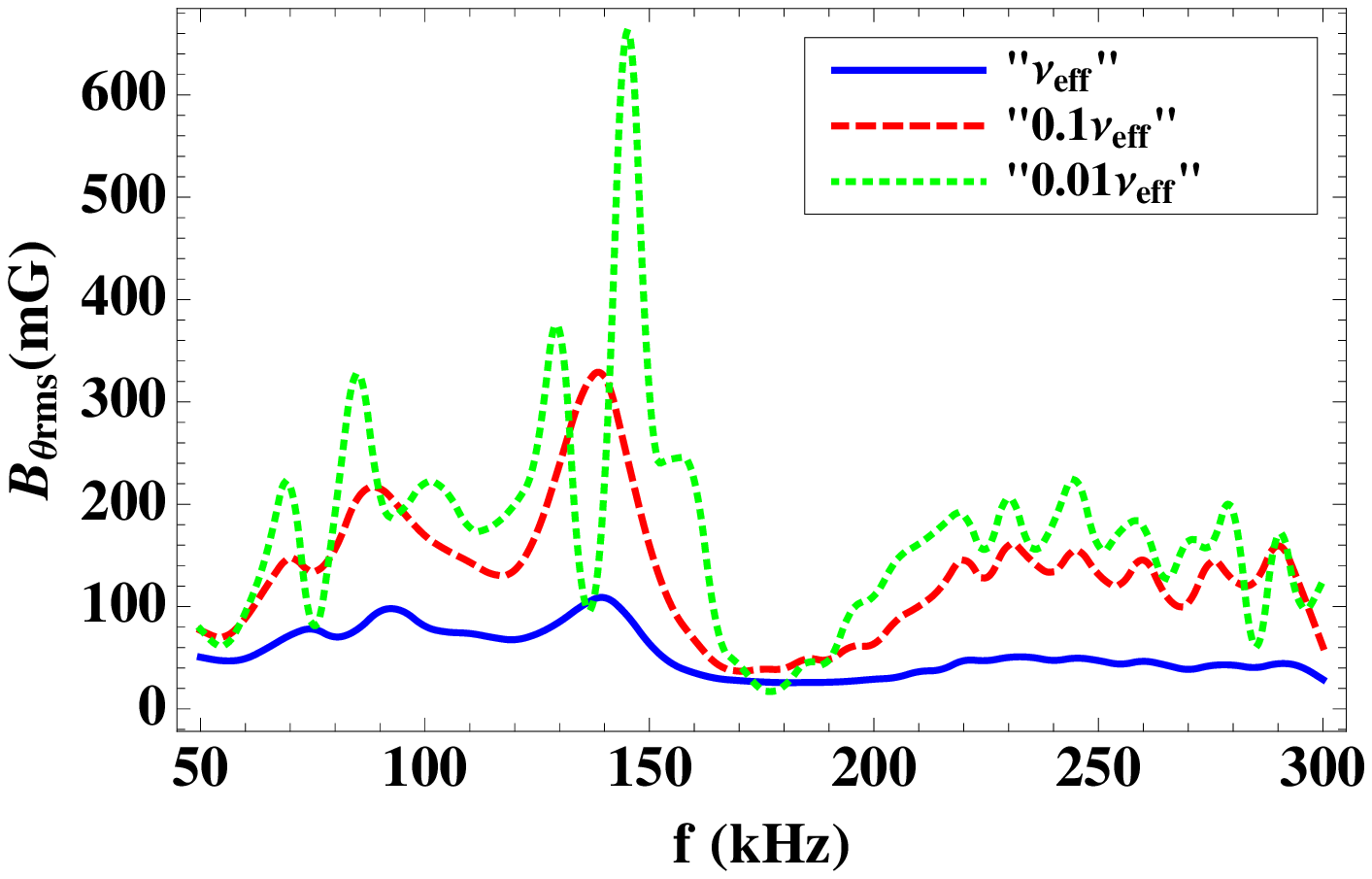}&\includegraphics[width=0.505\textwidth,angle=0]{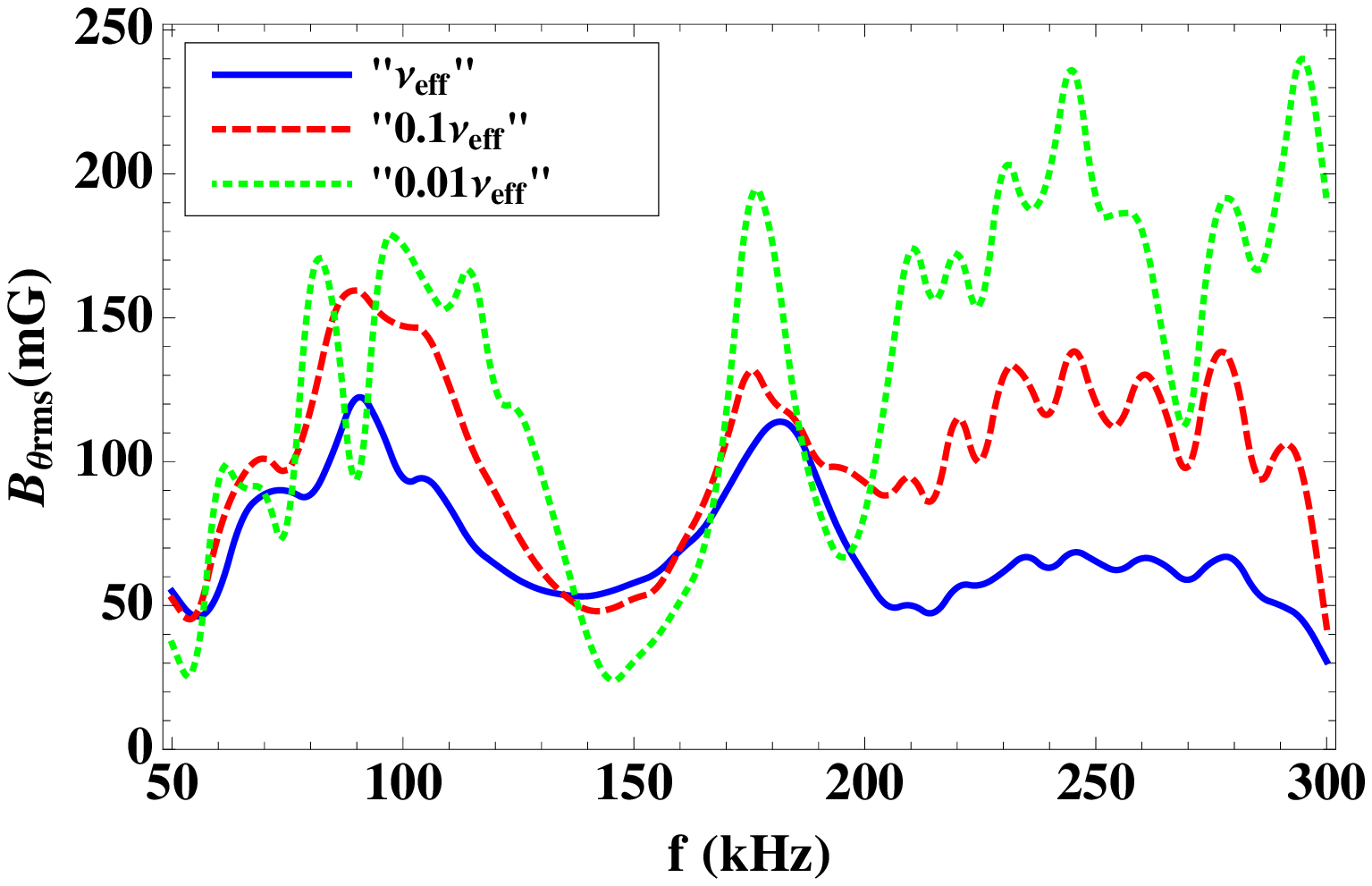}
\end{array}$
\end{center}
\caption{Variations of the wave magnitudes (rms) with frequency for the odd-parity AGE at $z=23.6975$~m and $r=0.0285$~m (a), and for the even-parity AGE at $z=25.5125$~m and $r=0.0371$~m (b). A significant drop of mode strength is clearly seen as the total effective collision frequency is made higher.}
\label{fg13}
\end{figure}

This section aims to guide the experimental implementation of gap eigenmode based on the realistic conditions of LAPD. First, the field strength will be decreased to the present capability of LAPD, and magnetic field of $B_0(z)=0.2+0.1\cos[2\pi(z-33.68)/3.63]$ (maximum $0.3$~T) with absorbing beach is thereby employed. Second, the plasma density will be reduced to $2.56\times10^{18}~\textrm{m}^{-3}$ (less than $2\times10^{19}~\textrm{m}^{-3}$) to maintain the same phase velocity of Alfv\'{e}n waves, allowing comparison with previous sections in the same frequency range. The computed gap eigenmodes are shown in Fig.~\ref{fg12}, which peak at $f=140$~kHz and $f=180$~kHz for odd-parity and even-parity modes respectively. 
The employed configurations of defective magnetic field are the same to those shown in Fig.~\ref{fg11} except the level of strength which is $6$ times less here. It can be seen that both the odd-parity and even-parity gap eigenmodes are clearly visible inside the spectral gap, although their decay lengths are much shorter than those shown in Fig.~\ref{fg10}. This states that for the present conditions of LAPD, the gap eigenmode of SAW can be formed immediately once the mirror number is doubled. Moreover, the comparison between Fig.~\ref{fg10} and Fig.~\ref{fg12} implies that stronger magnetic field makes it easier for the experimental observation. Although these gap eigenmodes are still visible for original Coulomb collisions and electron Landau damping, the total effective collision frequency ($\nu_{\textrm{eff}}=\nu_{ei}+\nu_{e\textrm{-Landau}}$) has been multiplied by $0.1$ for clear formation in Fig.~\ref{fg10} and Fig.~\ref{fg12}. The effect of this collision frequency on gap eigenmode strength is shown in Fig.~\ref{fg13}. 
As discussed in our previous study\cite{Chang:2013aa}, the resonant peak decreases and broadens for larger values of $\nu_{\textrm{eff}}$, but it is still clearly visible even at the highest collision frequency.

\subsection{Reduced number of magnetic mirrors}
\begin{figure}[ht]
\begin{center}$
\begin{array}{ll}
(a)&(b)\\
\includegraphics[width=0.495\textwidth,angle=0]{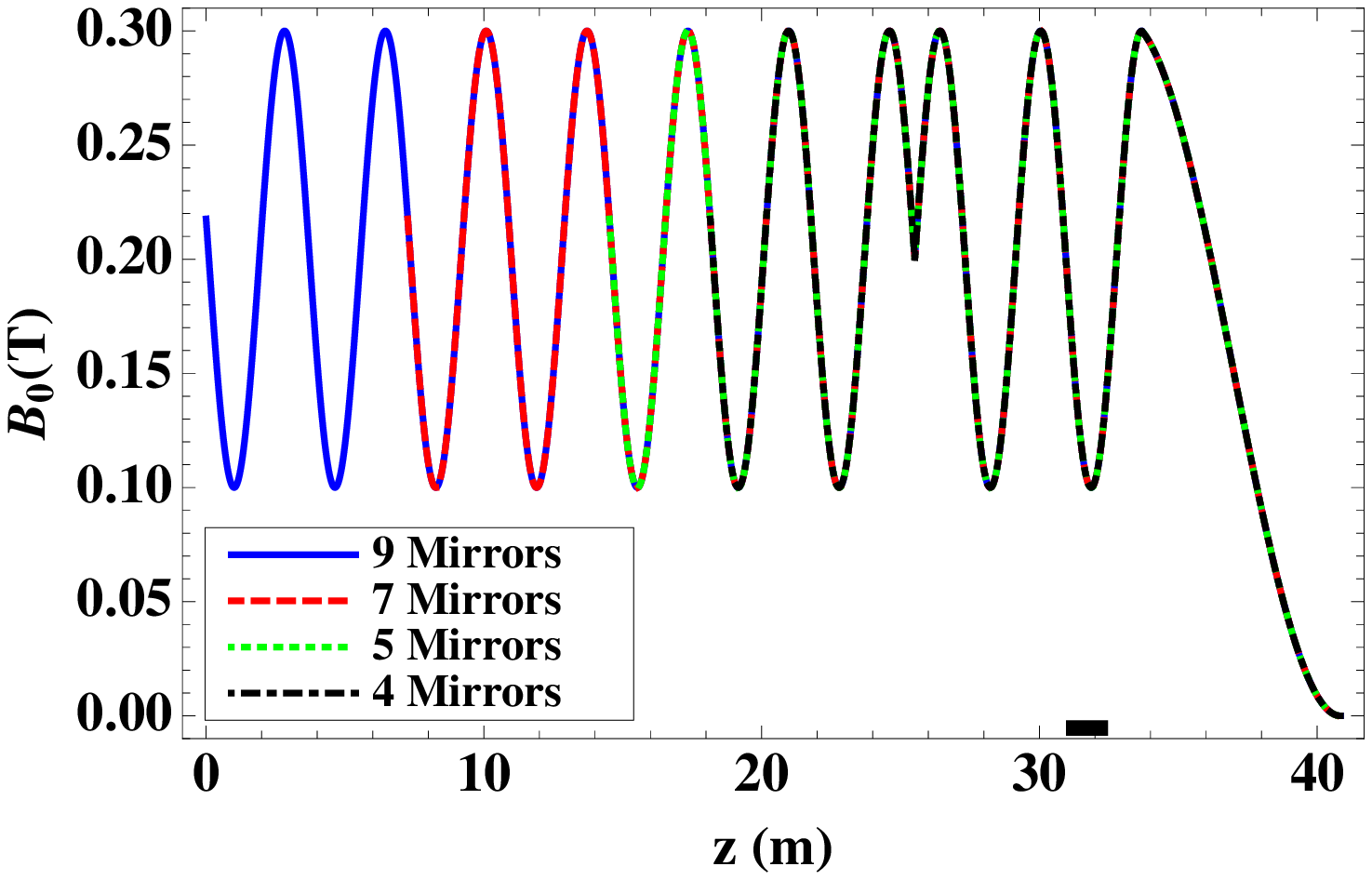}&\includegraphics[width=0.49\textwidth,angle=0]{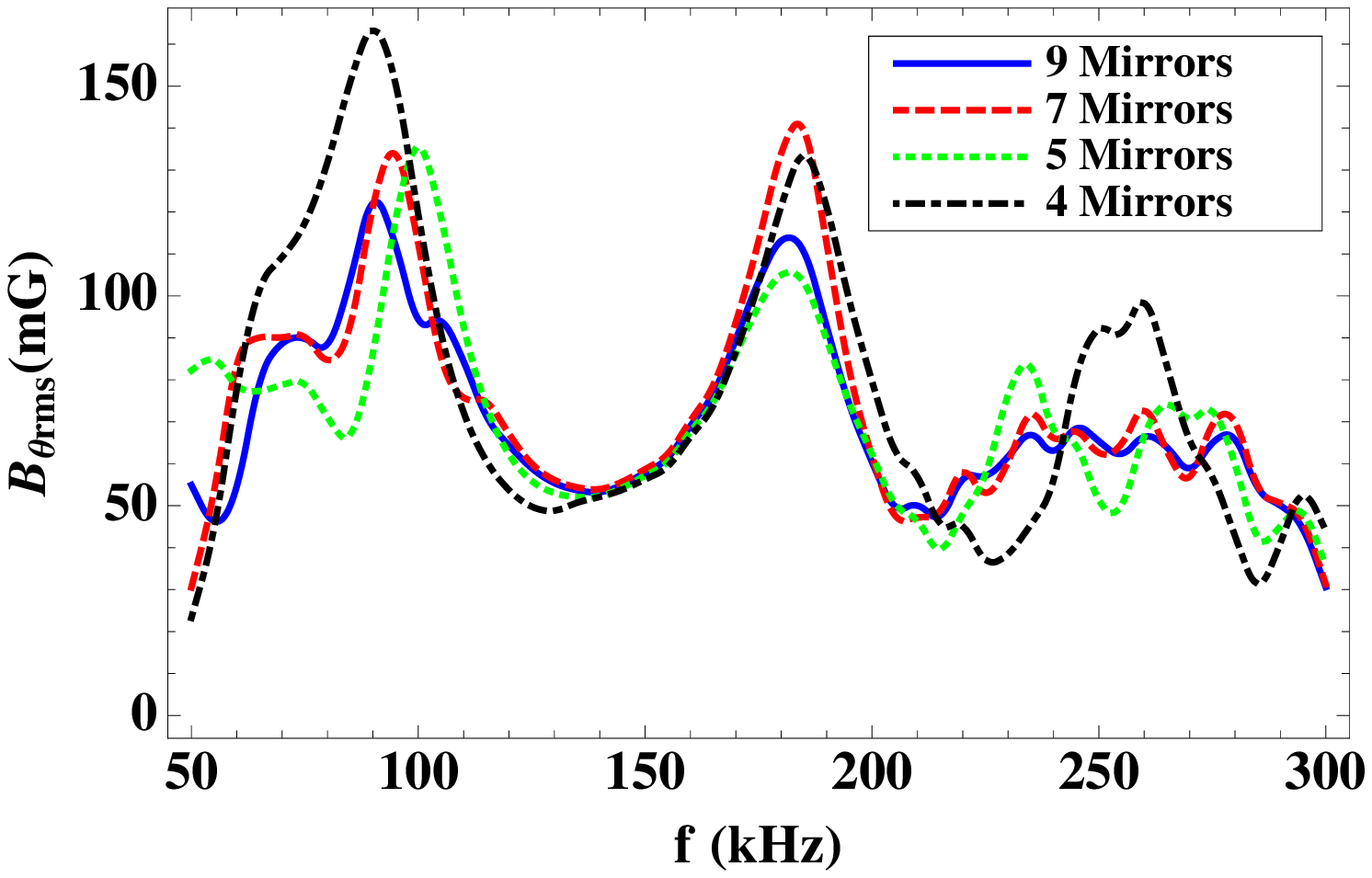}
\end{array}$
\end{center}
\caption{Configurations of defective magnetic field with different numbers of mirrors (a) and the resulted variations of peak wave magnitude (rms) with frequency. These peak magnitudes are measured at $r=0.0371$~m and axially $1.4548$~m away from the defect location. The black bar labels the external driving antenna.}
\label{fg14}
\end{figure}

\begin{figure}[ht]
\begin{center}$
\begin{array}{ll}
(a_1)&(b_1)\\
\includegraphics[width=0.48\textwidth,angle=0]{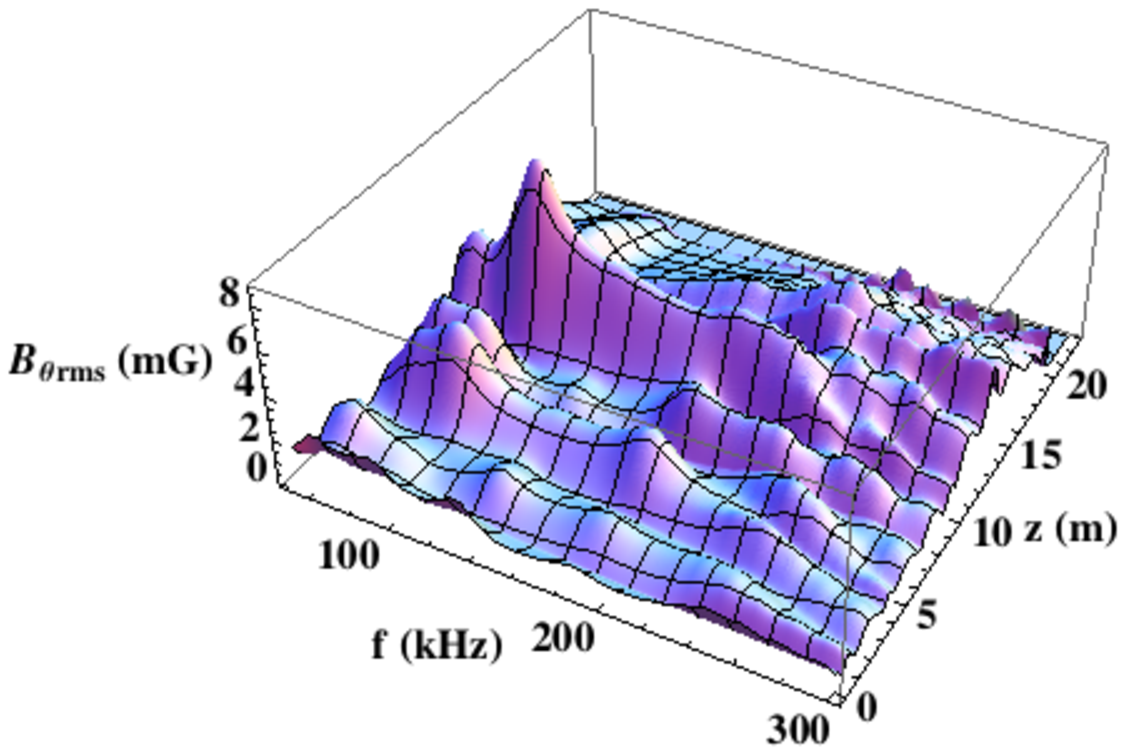}&\includegraphics[width=0.48\textwidth,angle=0]{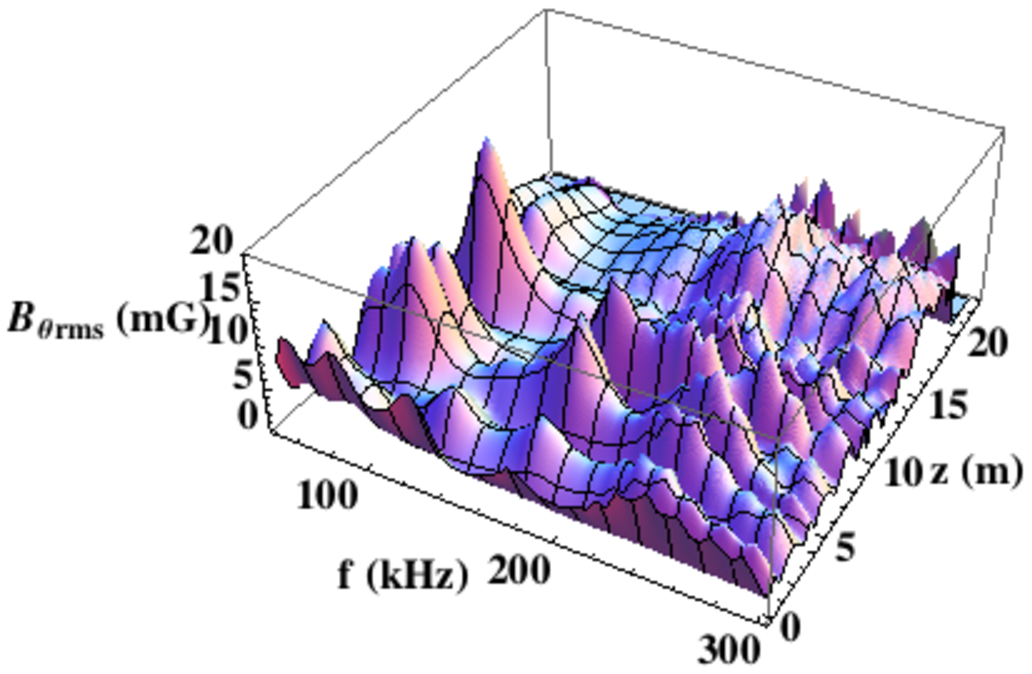}\\
(a_2)&(b_2)\\
\includegraphics[width=0.48\textwidth,angle=0]{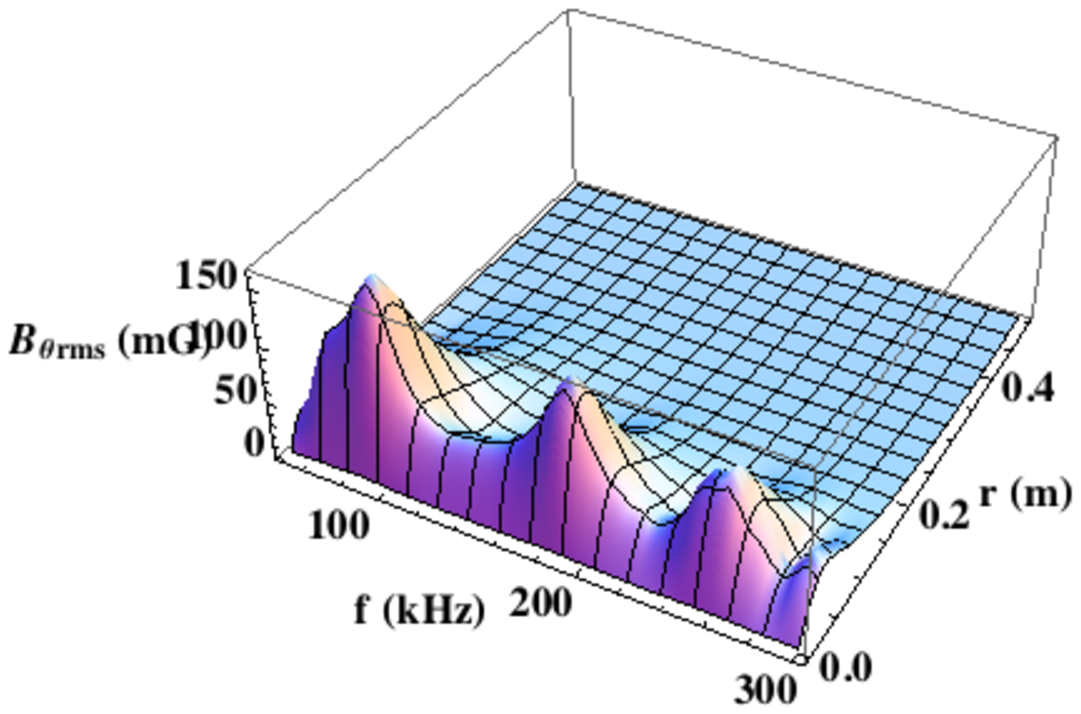}&\includegraphics[width=0.48\textwidth,angle=0]{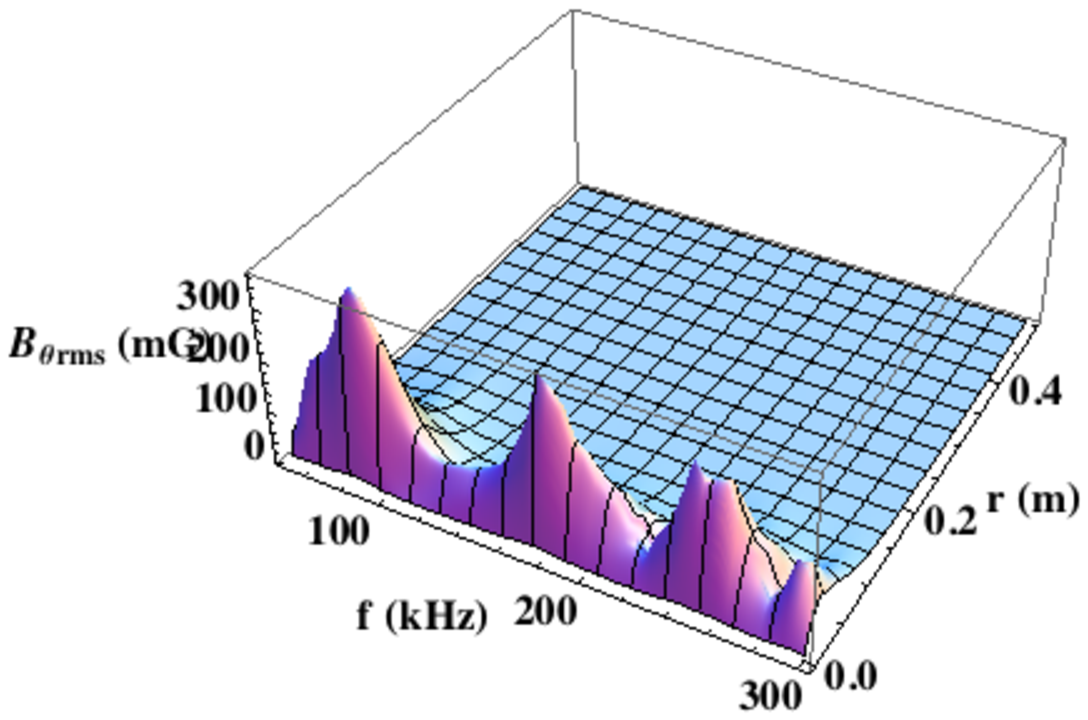}
\end{array}$
\end{center}
\caption{Surface plots of AGE in the $(f; z)$ and $(f; r)$ spaces for $4$ magnetic mirrors. For left figures ($a_1$ and $a_2$), the employed collisionality consists of original Coulomb collisions and electron Landau damping, whereas it is decreased by a factor of $10$ for right figures ($b_1$ and $b_2$).}
\label{fg15}
\end{figure}

Now we shall remove the last limitation for immediate experiment on LAPD by decreasing the number of magnetic mirrors. The even-parity gap eigenmode for original Coulomb collisions and electron Landau damping will be employed. The odd-parity gap eigenmode exhibits similar features and are thereby omitted. The constructed profiles of defective magnetic field with different numbers of mirrors are shown in Fig.~\ref{fg14}(a), and the resulted variations of peak wave magnitude (rms) with frequency are shown in Fig.~\ref{fg14}(b). These peak magnitudes are measured at $r=0.0371$~m and axially $1.4548$~m away from the defect location. It can be seen that the gap eigenmode is clearly formed even for the least number of magnetic mirrors, namely $4$ on LAPD. The reduction of magnetic mirrors has negligible effect on the frequency and width of the gap eigenmode.
The three-dimensional surface visualizations of this gap eigenmode for $4$ magnetic mirrors are shown in both $(f; z)$ and $(f; r)$ spaces in Fig.~\ref{fg15}. For left figures, the employed collisionality consists of original Coulomb collisions and electron Landau damping, whereas it is decreased by a factor of $10$ for right figures as done in Fig.~\ref{fg10} and Fig.~\ref{fg12}. Although the spectral gaps become less clear, compared with those in Fig.~\ref{fg12}($b_1$) and Fig.~\ref{fg12}($b_2$), the formed gap eigenmodes are still visible, especially in the radial direction. This strongly motivates the experimental implementation of this gap eigenmode on LAPD. 

\section{Discussion and Conclusion}\label{dcl}
\begin{figure}[ht]
\begin{center}$
\begin{array}{ll}
(a)&(b)\\
\includegraphics[width=0.5\textwidth,angle=0]{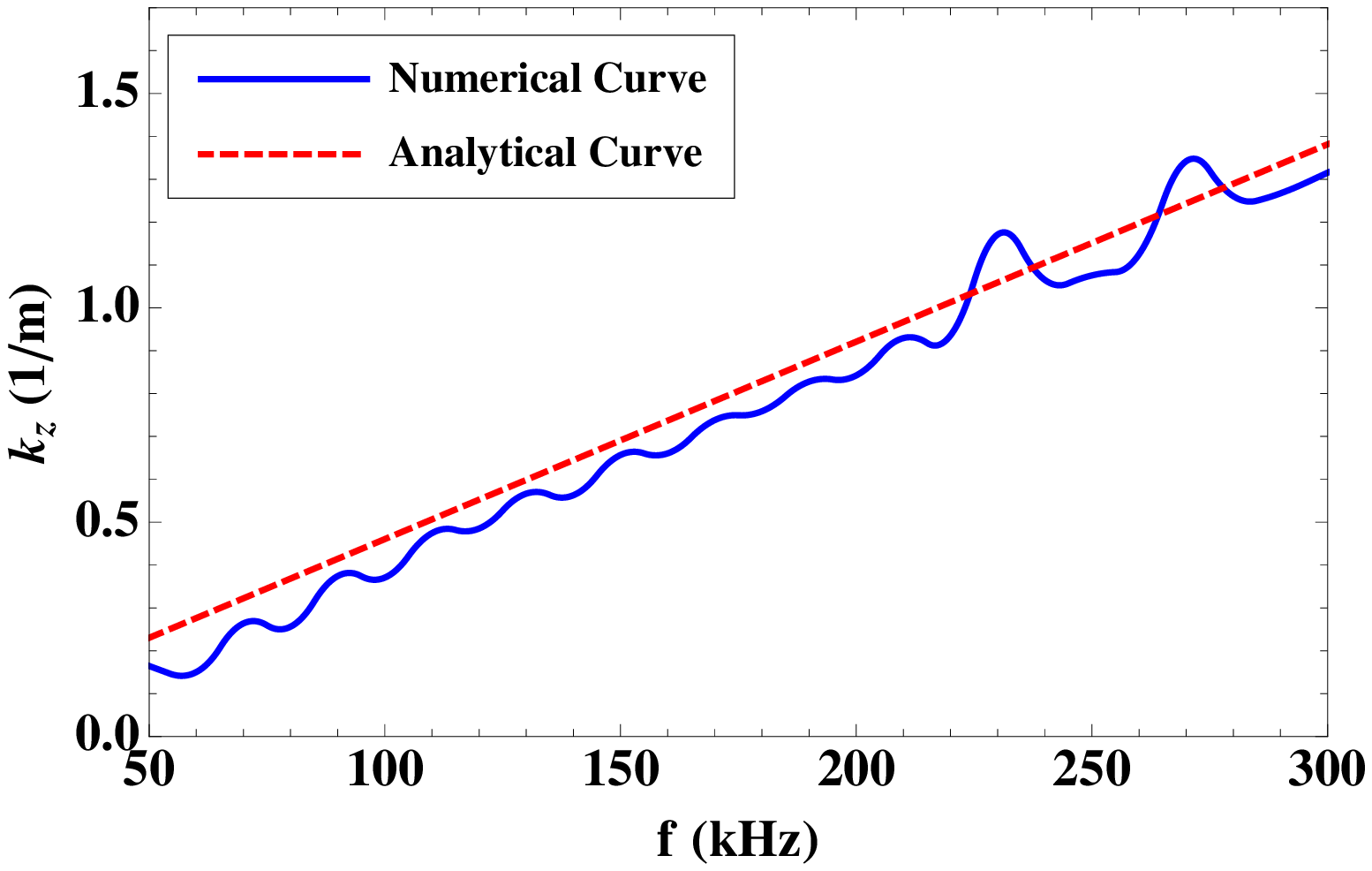}&\includegraphics[width=0.5\textwidth,angle=0]{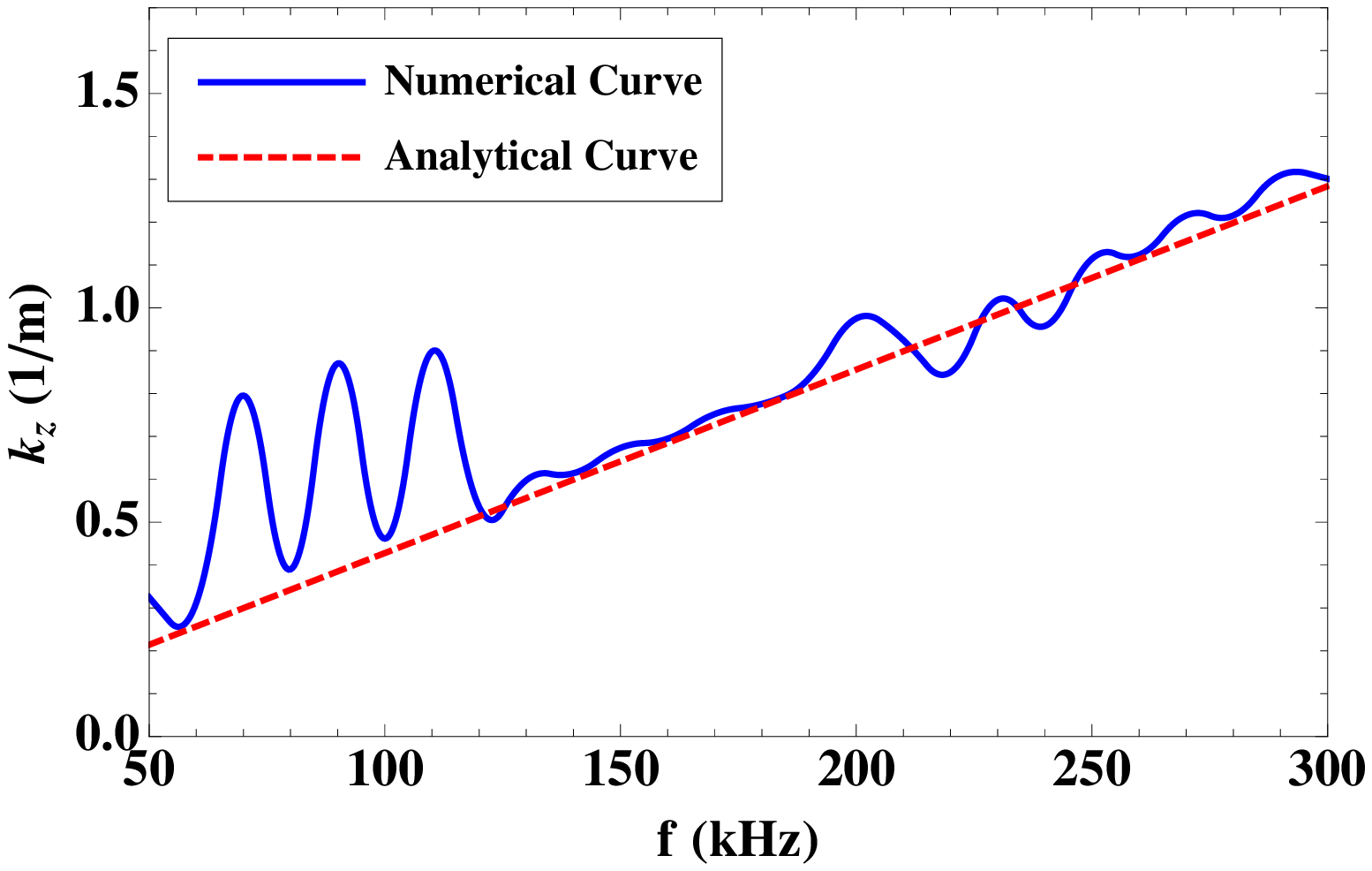}
\end{array}$
\end{center}
\caption{Numerical and analytical dispersion curves of Alfv\'{e}n waves on axis (a, $r=0$~m) and off axis (b, $r=0.12$~m).}
\label{fg16}
\end{figure}

\begin{figure}[ht]
\begin{center}
\includegraphics[width=0.5\textwidth,angle=0]{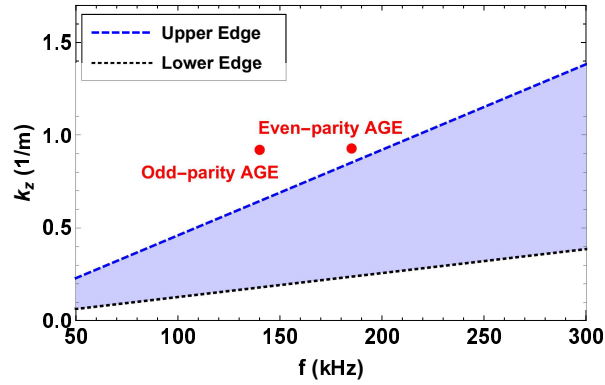}
\end{center}
\caption{Continuum damping region and formed AGEs.}
\label{fg17}
\end{figure}

To check whether the gap eigenmode shown in Fig.~\ref{fg10} is shear Alfv\'{e}n mode or compressional Alfv\'{e}n mode, we run EMS for various driving frequencies for uniform static magnetic field, and calculate the dominant axial wavenumber $k_z$ via Fourier decomposition. The computed dispersion curves at $r=0$~m and $r=0.12$~m are given in Fig.~\ref{fg16}, together with their comparisons with analytical theory. We can see that numerical and analytical curves agree well (wiggles are caused by reflections from endplate and can be smoothed when the endplate is moved further away). Moreover, the considered frequency range $50-300$~kHz is far below the ion cyclotron frequency $f_{ci}=4.57$~MHz. Therefore, it is reasonable to claim that the formed gap eigenmode here belongs to shear Alfv\'{e}n branch. 
To check whether the continuum damping resonance of Alfv\'{e}n waves effects the gap eigenmode formation, we compute the continuum damping region which is enclosed by Alfv\'{e}nic dispersion curves on plasma core and at plasma edge. Figure~\ref{fg17} illustrates the computed continuum damping region, which is enclosed by dispersion relations of $k_z=2\pi f/v_A$ on axis and at plasma edge respectively, and the locations of peak gap eigenmodes in Fig.~\ref{fg10}. It is clear that both odd-parity ($140$~kHz) and even-parity ($185$~kHz) gap eigenmodes lay outside of the continuum damping region thereby are unaffected by continuum resonance. 

In summary, to guide the experimental implementation of the gap eigenmode of SAW on LAPD, this work computes the gap eigenmode using exact parameters of LAPD. We start from the reproduction of wave field for previous experiment, and find that previously observed spectral gap is not global but an axially local result. By increasing the number of magnetic mirrors and static magnetic field strength, we obtain a clear and global spectral gap in both axial and radial directions. After introducing local defects, two types of gap eigenmodes are formed inside the spectral gap, namely odd-parity and even-parity. These gap eigenmodes are standing waves localized around the defects, and have wavelength twice the periodicity of external magnetic mirror, which is consistent with Bragg's law and previous studies\cite{Chang:2013aa, Chang:2014aa, Chang:2016aa}. These gap eigenmodes can be successfully formed even for the field strength, plasma density and number of magnetic mirrors that are achievable on LAPD at its present status. This preliminary computation strongly motivates the experimental observation of the gap eigenmode of SAW on LAPD and other relevant linear plasma devices. 

\ack
Many inspiring and helpful discussions with Boris Breizman and Matthew Hole are deeply appreciated. This work is supported by various funding sources: National Natural Science Foundation of China (11405271), China Postdoctoral Science Foundation (2017M612901), Chongqing Science and Technology Commission (cstc2017jcyjAX0047), Chongqing Postdoctoral Special Foundation (Xm2017109), Fundamental Research Funds for Central Universities (YJ201796), Pre-research of Key Laboratory Fund for Equipment (61422070306), and Laboratory of Advanced Space Propulsion (LabASP-2017-10).

\section*{References}
\bibliographystyle{unsrt}

\end{document}